\documentclass[pra,twocolumn,tightenlines,10pt,amsfonts,amssymb,balancelastpage]{revtex4-1}%
\usepackage[utf8x]{inputenc}

\usepackage{graphics}
\usepackage{epsfig}
\usepackage{subfigure}
\usepackage{rotating}

\usepackage{amsfonts}
\usepackage{amsmath}
\usepackage{amssymb}
\usepackage{placeins}

\usepackage{pgfplots}
\usepackage{tikz}
\usetikzlibrary{arrows,decorations.markings}

\usepackage{soul}
\usepackage{fixmath}

\begin{document}

\title{A modal approach to quantum temporal imaging}
\author{G. Patera$^1$, M. Allgaier$^2$, D. B. Horoshko$^3$, M. I. Kolobov$^1$, and C. Silberhorn$^4$}
\address{$^1$Univ. Lille, CNRS, UMR 8523 - PhLAM - Physique des Lasers Atomes et Mol\'{e}cules, F-59000 Lille, France\\
$^2$ Department of Physics and Oregon Center for Optical, Molecular, and Quantum Science, University of
Oregon, Eugene, OR 97403, USA\\
$^3$ B. I. Stepanov Institute of Physics, NASB, Nezavisimosti Ave. 68, Minsk 220072 Belarus\\
$^4$ Integrated Quantum Optics Group, Applied Physics, University of Paderborn, 33098 Paderborn, Germany}

\begin{abstract}
We consider the problem of quantum temporal imaging in the case where the time lens is implemented by a Sum Frequency Generation nonlinear process, in particular when the device is operated close to $100\%$ conversion efficiency. In the general case where the time lens also presents a finite aperture and a non-perfect phase-matching the relevant figures of merit, as for example the temporal resolution, do not have an explicit expression. As a consequence, the performances of imaging scheme are difficult to assess. Having a reliable estimation of these figures of merits is particularly important because they can be significantly different from the regime of low conversion efficiency usually considered in classical temporal imaging. We show that this problem can be approached in terms of the eigenmodes of the imaging scheme and we show how its relevant figures of  merit can be extracted from the modal description of the imaging scheme. As a consequence we obtain criteria allowing to design imaging schemes with close to unity efficiencies.
\end{abstract}

\date{\today}
\maketitle

\section{Introduction}
Optical spectrotemporal degrees of freedom provide a reliable and flexible encoding for photonic non-classical states~\cite{Brecht2015}. Their noiseless manipulation plays an important role in the processing of quantum information and communication networks \cite{Raymer2012}.
In order to cope with the great heterogeneity of technologies that are used in a quantum network, it is important to optimize mode-matching between nodes. To this purpose a plethora of strategies have been proposed such as quantum translation of the carrier frequency~\cite{Tanzilli2005,McKinstrie2005,McGuinness2010,Karpinski2021}, waveform conversion~\cite{Kielpinski2011}, bandwidth compression or stretching~\cite{Lavoie2013,Karpinski2016,Allgaier2017,Wright2017,Sosnicki2020}, quantum pulse gating~\cite{Brecht2011,Eckstein2011} and shaping~\cite{Brecht2014,Manurkar2016}. 

Another approach to the manipulation of optical pulses is based on the formal space-time analogy between the propagation of a diffracting beam and that of a short pulse in dispersive media. For classical beams, this led to the formulation of temporal imaging~\cite{Kolner1994,Bennett2000a,Bennett2000b} and the demonstration of ultrafast waveform magnification and reversal~\cite{Bennett1994,Bennett1999}, compression~\cite{Foster2009} or ultrafast waveform detection~\cite{Foster2008,Tikan2018}. On the other side, quantum temporal imaging (QTI) aims at the manipulation of the time-frequency degrees of freedom of a quantum state without destroying it. This has been considered either for single photons~\cite{Zhu2013,Donohue2015,Karpinski2016,Mittal2017,Shi2020,Sosnicki2018} or for squeezed light~\cite{Patera2015,Shi2017,Patera2017,Shi2018,Patera2018}.

The key element of a temporal imaging system is the time lens, a device that imprints a quadratic temporal phase modulation on an input pulse as like a thin lens induces a quadratic phase modulation on a spatially extended input wavefront. Optical time lenses are presently based on electro-optical phase modulation~\cite{Kolner1989,Karpinski2016,Sosnicki2020}, sum-frequency generation (SFG)~\cite{Bennett1994,Bennett1999}, or four-wave mixing (FWM)~\cite{Foster2008,Foster2009,Okawachi2009,Kuzucu2009} and provide a temporal magnification up to 100 times. Time-lenses based on non-linear processes, as discussed in \cite{Zhu2013,Patera2015,Patera2017,Shi2017}, need to be operated in the regime close to 100\% conversion efficiency in order to process an input quantum state without destroying its quantum properties. This case has been considered in the limit of infinite aperture and perfect phase-matching \cite{Zhu2013,Donohue2015,Patera2015,Shi2017} and in the limit of finite aperture but still perfect phase-matching \cite{Patera2018,Shi2018}. Here we consider the most general case of QTI with finite aperture and in presence of non perfect phase-matching. This case has been considered in \cite{Bennett2000b} but only in the low conversion efficiency approximation, a regime suitable for classical protocols only. On the contrary, in the high conversion efficiency regime this problem does not admit a closed-form solution, thus an explicit expression for the impulse response function (IRF) is not possible. The IRF represents the response of the system to a point object and plays an important role for extracting the relevant figures of merit (as the temporal resolution) and for assessing the system performances. In this work we show that the problem can be approached in terms of the eigenmodes of the imaging scheme and that its relevant figures of merit can be extracted from the modal description even if the IRF does not have an explicit expression. Our approach also allows to make clear the multimode nature of QTI, a character that is important for processing an image without distortions, thus enabling the design of schemes with close to unit efficiencies.
It is interesting to remark that despite being implemented by the same non-linear process of SFG-based QTI, quantum pulse gates and shapers~\cite{Brecht2011,Eckstein2011,Brecht2014,Manurkar2016} are rather designed for single mode operation.
\section{The formalism}
\begin{figure}[t!]
\centering
\subfigure[\label{fig:imag scheme}]
{\resizebox{1.\columnwidth}{!}{\includegraphics{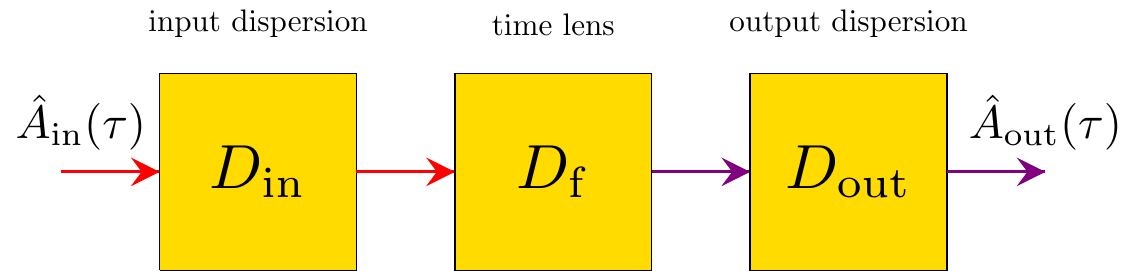}}}
\\
\subfigure[\label{fig:TL scheme}]
{\resizebox{0.5\columnwidth}{!}{\includegraphics{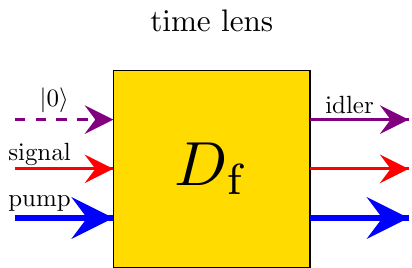}}}
\caption{(a) The imaging scheme: input dispersive line
         of total GGD $D_{\mathrm{in}}$ is followed by a time lens of equivalent GDD $D_{\mathrm{f}}$ 
				 and by an output dispersive line of total GDD $D_{\mathrm{out}}$.
         The input field $\hat{A}_{\mathrm{in}}(\tau)$ is imaged at the output
	       on the field $\hat{A}_{\mathrm{out}}(\tau)$. (b) SFG-based time lens scheme: the signal field (red) is mixed together with a strong classical pump (blue) in order to produce the idler field (purple).}
\end{figure}

In this paper we consider the simplest imaging scheme that can be realized with a single time lens as depicted in figure \ref{fig:imag scheme}: it consists of a first dispersive medium followed by a time lens then followed by a second dispersive medium. This scheme is the temporal equivalent of a thin lens that performs the imaging of a spatially extended object. In the following we will refer to the first (second) medium as the \textit{input} (\textit{output}) dispersive medium. Without loss of generality, we will focus on a SFG-based time lens even if our approach remains valid also for FWM-based time lenses. The SFG process (see figure \ref{fig:TL scheme}) is mediated by a short chirped pump pulse of carrier frequency $\omega_{\mathrm{p}}$. Hence a pulse at signal frequency $\omega_{\mathrm{s}}$ after a dispersive propagation through the input medium is up-converted in a non-linear crystal to a new pulse at the idler frequency $\omega_{\mathrm{i}}$ such that $\omega_{\mathrm{s}}+\omega_{\mathrm{p}}=\omega_{\mathrm{i}}$.
Finally the idler pulse is dispersed through the output medium.
%
%
%
In the plane wave and quasi-monochromatic approximation we write the positive-frequency electric field operator as
\begin{align}
\hat{E}^{(+)}_m(t,z)&=
\mathcal{E}_m\,
\mathrm{e}^{\mathrm{i}\left(k_m z-\omega_m t\right)}
\hat{A}_m(t,z),
\\
\hat{A}_m(t,z)&=
\int_{-\infty}^{+\infty}
\frac{\mathrm{d}\Omega}{2\pi}\,
\mathrm{e}^{-\mathrm{i}\Omega t}
\mathrm{e}^{\mathrm{i}\,\delta_m(\Omega)z}\hat{a}_m(\Omega,z)
\label{A}
\end{align}
where the index $m=\{\mathrm{s},\mathrm{i},\mathrm{p}\}$ identifies the signal, idler or pumps waves respectively, $\mathcal{E}_m$ is the single photon amplitude, $k_m=k(\omega_m)$, $\delta_m(\Omega)=k(\omega_m+\Omega)-k_m$.

It is convenient to define a travelling-wave frame of reference $(\tau,z)$ propagating with the wave
at pulse group velocity where the \textit{delayed time} $\tau$ is defined as
\begin{align*}
\tau&=t-\beta^{(1)}_m z
\end{align*}
with $\beta^{(1)}_m=(\mathrm{d} k/\mathrm{d}\omega)_{\omega_m}$ the inverse of group velocity.
In this frame of reference we have
\begin{align}
\hat{A}_m(\tau,z)&=
\int_{-\infty}^{+\infty}
\frac{\mathrm{d}\Omega}{2\pi}\,
\mathrm{e}^{-\mathrm{i}\Omega \tau}
\mathrm{e}^{\mathrm{i}\,\zeta_m(\Omega)z}
\hat{a}_m(\Omega,z)
\label{A tx}
\end{align}
where $\zeta_m(\Omega)=\delta_m(\Omega)-\beta^{(1)}_m\Omega$.

\subsection{Linear Dispersion}
In the quasi-monochromatic approximation, the evolution of the $\hat{A}_m(\tau,z)$ 
fields through a dispersive medium of length $L$ is given by
\begin{align}
\hat{A}_m(\tau,L)=
\int_{-\infty}^{+\infty}
\frac{\mathrm{d}\Omega}{2\pi}\,
\mathrm{e}^{-\mathrm{i}\Omega \tau}\,
\mathcal{G}(\Omega)\hat{a}_m(\Omega,0)
\label{disp evo}
\end{align}
where 
\begin{align}
\mathcal{G}(\Omega)=\mathrm{e}^{\frac{\mathrm{i}}{2}D_m\Omega^2},
\end{align}
$\beta^{(2)}_m=(\mathrm{d}^2 k/\mathrm{d}\omega^2)_{\omega_m}$ is the Group Velocity Dispersion (GVD) and
$D_m=\beta^{(2)}_m L$ is the Group Delay Dispersion (GDD) accumulated by the wave during its dispersive propagation through the medium.
Also eq. \eqref{disp evo} can rephrased as
\begin{align}
\hat{A}_m(\tau,L)=
\int_{-\infty}^{+\infty}
\mathrm{d}\tau'\,
\mathrm{e}^{\mathrm{i}\Omega \tau}\,
G(\tau-\tau')\hat{A}_m(\tau',0)
\label{disp evo II}
\end{align}
with
\begin{equation}
G(\tau)=
\frac{\mathrm{e}^{-\mathrm{i}\tau^2/2 D_m}}{\sqrt{-\mathrm{i}2\pi D_m}}
\end{equation}
Notice that eqs. \eqref{disp evo} and \eqref{disp evo II} induce a unitary transformation on the field operators, thus
preserving their commutators.


\subsection{Time lens}\label{Timelens}
For a SFG-based time lens, the object pulse enters in the nonlinear medium through the input signal channel and is up-converted in the output idler mode (see figure~\ref{fig:TL scheme}). The input idler mode is in vacuum state and the process is mediated by a strong undepleted pump. The evolution equations for signal ($\hat{a}_{\mathrm{s}}$) and idler ($\hat{a}_{\mathrm{i}}$) fields through the nonlinear crystal, in the plane-wave and undepleted pump approximations, are given by
\begin{align}
\frac{\partial}{\partial \xi}\hat{a}_{\mathrm{s}}(\Omega,\xi)&=
g\int\mathrm{d}\Omega'
f^{*}(\Omega',\Omega,\xi)
\hat{a}_{\mathrm{i}}(\Omega',\xi),\label{evo as}
\\
\frac{\partial}{\partial \xi}\hat{a}_{\mathrm{i}}(\Omega,\xi)&=
-g\int\mathrm{d}\Omega'
f(\Omega,\Omega',\xi)
\hat{a}_{\mathrm{s}}(\Omega',\xi),\label{evo ai}
\end{align}
where
\begin{equation}\label{coupling}
f(\Omega,\Omega',\xi)=\alpha_{\mathrm{p}}(\Omega-\Omega')
\mathrm{e}^{-\mathrm{i}\Delta(\Omega,\Omega')\xi}
\end{equation}
$\alpha_{\mathrm{p}}(\Omega)$ are the Fourier amplitudes of the pump classical field at the
input of the crystal (since we assumed it does not evolve along $z$), $g$ is a parameter
proportional to the effective nonlinear coupling inside the crystal and the function
$\Delta(\Omega,\Omega')$ is the phase mismatch between signal, idler and pump spectral components defined as following
\begin{align}\label{mismatch}
\Delta(\Omega,\Omega')=
k_{\mathrm{i},z}(\Omega,\mathbf{q}_{\mathrm{i}})
-k_{\mathrm{s},z}(\Omega',\mathbf{q}_{\mathrm{s}})-k_{\mathrm{p},z}(\Omega-\Omega',\mathbf{q}_{\mathrm{p}})
\end{align}
where we allow for the general case of plane waves non-collinear with the longitudinal axis $z$,
presenting a transverse component of the wave-vector $\mathbf{q}_{\mathrm{s}}$, $\mathbf{q}_{\mathrm{i}}$ and $\mathbf{q}_{\mathrm{p}}$
for signal, idler and pump fields respectively such that
\begin{align}
k_{m,z}(\Omega,\mathbf{q}_{m})=\sqrt{k_m^2(\omega_m+\Omega)-\mathbf{q}_m^2}.
\end{align}
%
%

The solution of eqs. \eqref{evo as} and \eqref{evo ai} can be obtained as a linear symplectic integral transformation
for the fields operators using the Magnus perturbative approach. At the first order we have
\begin{align}
\left(
\begin{array}{c}
\hat{a}_{\mathrm{s}}(\Omega,l_c/2)
\\
\hat{a}_{\mathrm{i}}(\Omega,l_c/2)
\end{array}
\right)
&=
\mathrm{e}^{\int\mathrm{d}\Omega'\,M_1(\Omega,\Omega')}
\left(
\begin{array}{c}
\hat{a}_{\mathrm{s}}(\Omega',-l_c/2)
\\
\hat{a}_{\mathrm{i}}(\Omega',-l_c/2)
\end{array}
\right)
\label{sols}
\end{align}
where
\begin{align}
M_1(\Omega,\Omega')&=
\left(
\begin{array}{cc}
0 & g l_c K^*(\Omega',\Omega)
\\
-g l_c K(\Omega,\Omega') & 0
\end{array}
\right)
\end{align}
and
\begin{align}
K(\Omega,\Omega')&=
\int_{-l_c/2}^{+l_c/2}\mathrm{d}z\
f(\Omega,\Omega',z)=
\nonumber\\
&=
\alpha_{\mathrm{p}}(\Omega-\Omega')\; \mathrm{Sinc}\left(\Delta(\Omega,\Omega')\frac{l_c}{2}\right).
\label{K}
\end{align}
Notice that the function $K(\Omega,\Omega')$ is not symmetric with respect to the exchange $\Omega\leftrightarrow\Omega'$
since the function $\Delta(\Omega,\Omega')$ is not either.

Then by using the Singular Value Decomposition (SVD) of $K(\Omega,\Omega')$
that reads
\begin{align}
K(\Omega,\Omega')&=
\sum_m
\lambda_m
\psi_m(\Omega)\phi_m^*(\Omega')
\label{Kbis}
\end{align}
the solution \eqref{sols} can be put in the following form 
\begin{align}
\left(
\begin{array}{c}
\hat{a}_{\mathrm{s}}(\Omega,l_c/2)
\\
\hat{a}_{\mathrm{i}}(\Omega,l_c/2)
\end{array}
\right)
&=
\int\mathrm{d}\Omega'\,
B(\Omega,\Omega')
\left(
\begin{array}{c}
\hat{a}_{\mathrm{s}}(\Omega',-l_c/2)
\\
\hat{a}_{\mathrm{i}}(\Omega',-l_c/2)
\end{array}
\right)
\label{sols II}
\end{align}
where
\begin{align}
B(\Omega,\Omega')
&=
\left(
\begin{array}{cc}
U_{\mathrm{s}}(\Omega,\Omega') & V_{\mathrm{s}}(\Omega,\Omega')
\\
-V_{\mathrm{i}}(\Omega,\Omega') & U_{\mathrm{i}}(\Omega,\Omega')
\end{array}
\right)
\label{B}
\end{align}
and
\begin{align}
U_{\mathrm{s}}(\Omega,\Omega')&=
\sum_m \cos(g l_c \lambda_m)\phi_m(\Omega)\phi^*_m(\Omega'),
\label{Us}
\\
V_{\mathrm{s}}(\Omega,\Omega')&=
\sum_m \sin(g l_c \lambda_m)\phi_m(\Omega)\psi^*_m(\Omega'),
\label{Vs}
\\
U_{\mathrm{i}}(\Omega,\Omega')&=
\sum_m \cos(g l_c \lambda_m)\psi_m(\Omega)\psi^*_m(\Omega'),
\label{Ui}
\\
V_{\mathrm{i}}(\Omega,\Omega')&=
\sum_m \sin(g l_c \lambda_m)\psi_m(\Omega)\phi^*_m(\Omega').
\label{Vi}
\end{align}

Hence the transformation induced by the time lens does not have a closed-form expression, but it is given as an expansion in terms of the singular values $\lambda_m$ and eigenfunctions $\{\psi_m\}$ and $\{\phi_m\}$ of the problem.

We note that a 1st order Magnus perturbation theory is accurate up to $80\%$ conversion efficiencies and it introduces errors when attempting to discuss regimes much closes to $100\%$ conversion efficiency. In particular when the 1st order theory predicts a $100\%$ conversion efficiency, the exact model presents about an $80\%$ efficiency~\cite{Christ2013}. However, the modal method we are considering in this work is general and remains valid for any order of the Magnus expansion and we discuss the 1st order for sake of simplicity.

\subsubsection*{SFG configurations}\label{PM}
The SFG process can be configured in different ways according to the chosen parameters for the pump and phase-matching.

For what concerns the pump, a time lens is obtained when using a chirped Gaussian pulse. For this, a short Fourier-limited pulse of duration $\tau_{\mathrm{p}}$ and spectral bandwidth $\Delta_{\mathrm{p}}$ is dispersed through a medium of GDD $D_{\mathrm{p}}=\beta_{\mathrm{p}}^{(2)} L_{\mathrm{p}}$, with $L_{\mathrm{p}}$ the length and $\beta_{\mathrm{p}}^{(2)}$ the group velocity dispersion of the medium. After the propagation, in the Fraunhofer dispersion limit, the pulse is stretched to a duration $\tau'_{\mathrm{p}}\gg\tau_{\mathrm{p}}$: 
\begin{equation}
\tau'_{\mathrm{p}}=D_{\mathrm{p}}\Delta_{\mathrm{p}}.
\end{equation}
In the Fourier domain the chirped pulse is
\begin{align}
\alpha_{\mathrm{p}}(\Omega)
&=
A_{\mathrm{p}}\,\mathrm{e}^{-\frac{1}{2}\Omega^2/\Delta_{\mathrm{p}}^2}\,
\mathrm{e}^{\frac{\mathrm{i}}{2}D_{\mathrm{f}}\Omega^2},
\label{chirped Gauss pump}
\end{align}
where we define $D_{\mathrm{f}}=-D_{\mathrm{p}}$ is the \textit{focal GDD} of the time lens.

For what concerns the phase-matching, we can Taylor expand expression.~\eqref{mismatch} up to first order in $\Omega$ and $\Omega'$
\begin{equation}\label{mism 1st order}
\Delta(\Omega,\Omega')\frac{l_c}{2}\approx
\Delta_0 \frac{l_c}{2}
+
(k'_{\mathrm{i}}-k'_{\mathrm{p}})\frac{l_c}{2} \Omega+
(k'_{\mathrm{p}}-k'_{\mathrm{s}})\frac{l_c}{2}\Omega'
\end{equation}
where $\Delta_0=k_{\mathrm{i}}-k_{\mathrm{s}}-k_{\mathrm{p}}$ and $k'_m$ are, respectively, the phase-mismatch and the group velocity at the carrier frequency $\omega_{m}$, for $m=\{\mathrm{s},\mathrm{i},\mathrm{p}\}$. We restrict our treatment to processes that are perfectly phase matched at the central frequencies of the three waves so that $\Delta_0 = 0$.

\begin{figure*}[!t]
\centering
\subfigure[\label{fig:K1 module}]{\resizebox{0.45\textwidth}{!}{\includegraphics{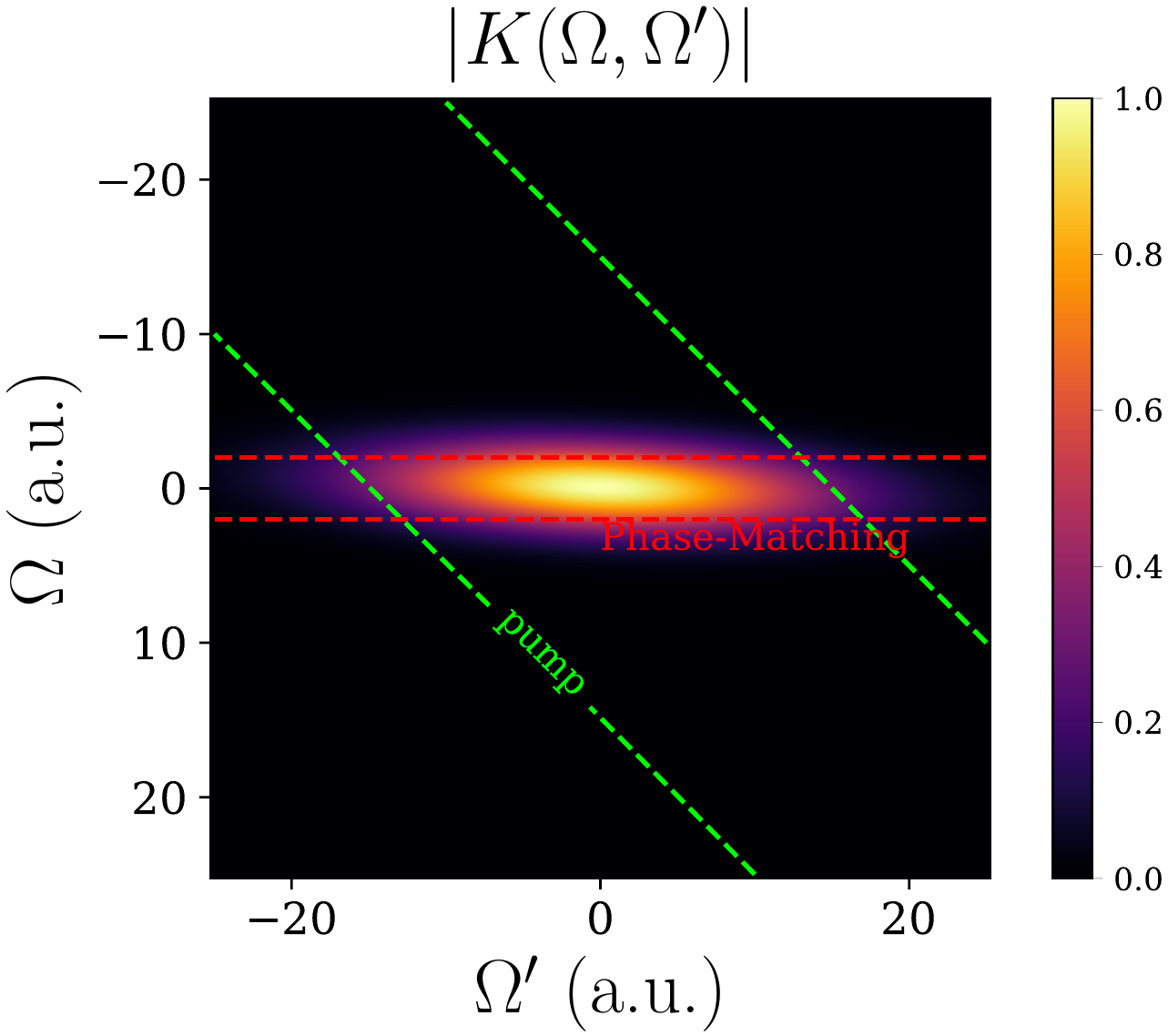}}}
\quad
\subfigure[\label{fig:K1 phase}]{\resizebox{0.45\textwidth}{!}{\includegraphics{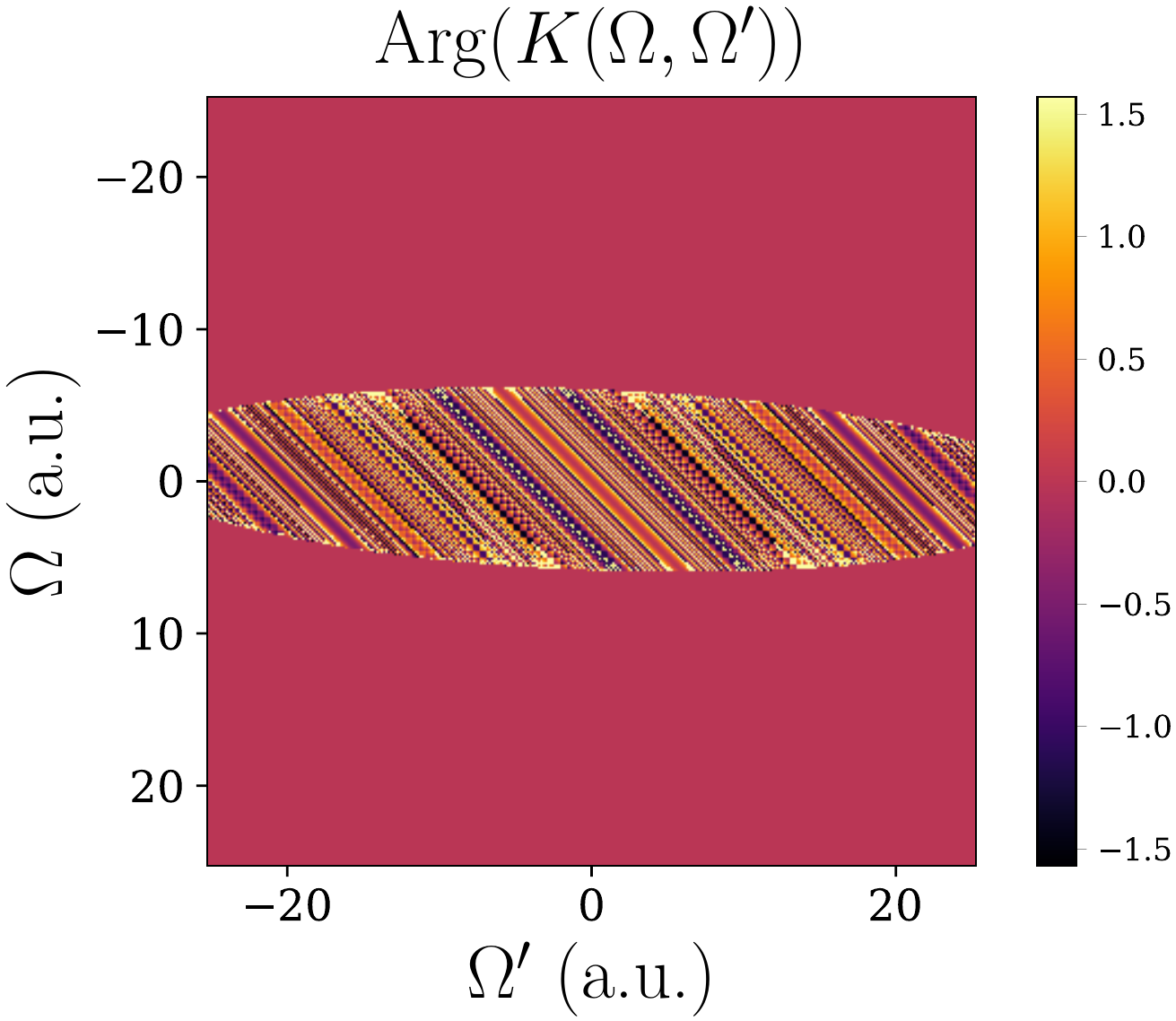}}} \\
\caption{(a) module and (b) phase of $K(\Omega,\Omega')$ as described in eq. \eqref{K1}. The parameters
are: $\Delta_h=100$ a.u., $\Delta_v=2$ a.u., $\Delta_{\mathrm{p}}=10$ a.u., $D_{\mathrm{f}}=5$ a.u.}
\label{fig:K1}
\end{figure*}
We can, then, distinguish three configurations that are qualitatively distinct:
\begin{enumerate}
    \item[(i)] Ideal: in this case not only the phase matching is perfect at carriers but also for every $\Omega$ and $\Omega'$ so that $\Delta(\Omega,\Omega')l_c/2=0$. This condition would require to simultaneously satisfy $(k'_{\mathrm{i}}-k'_{\mathrm{p}})l_c=0$ and $(k'_{\mathrm{p}}-k'_{\mathrm{s}})l_c=0$. The pump is assumed to be infinitely long;
    \item[(ii)] Perfect phase-matching and finite aperture: in this case the aperture of the time lens is determined only by the (finite) pump duration $\tau'_{\mathrm{p}}$ and it corresponds to the physical situation where the temporal walk-off between pump and the signal and idler is much smaller than the inverse of the pump bandwidth $\Delta_{\mathrm{p}}^{-1}=\tau_{\mathrm{p}}$. The conditions to be satisfied are now: $(k'_{\mathrm{i}}-k'_{\mathrm{p}})l_c\gg\tau_{\mathrm{p}}$ and $(k'_{\mathrm{p}}-k'_{\mathrm{s}})l_c\gg\tau_{\mathrm{p}}$. From an experimental point of view these conditions could be implemented by using symmetric group velocity matching $k'_{\mathrm{i}}-k'_{\mathrm{p}}=k'_{\mathrm{p}}-k'_{\mathrm{s}}$ (this condition is also known, in the case of parametric-down conversion as ``extended phase-matching"~\cite{Giovannetti2002}). Notice, however, that it is challenging to assure for the corresponding temporal walk-off to be larger than the reciprocal of the pump bandwidth. When these conditions are not respected, the temporal aperture is determined not only by the pump duration but also by the spectral filtering induced by the phase-matching. As a consequence the time aperture would be smaller than that determined by the pump only. From a classical point of view, one could tune the other free parameter, the crystal length $l_c$, to be sufficiently small in order to satisfy those conditions. On the other hand, in quantum regime, a small $l_c$ would reduce the conversion efficiency that should be compensated by higher pump intensities. A different implementation could be via asymmetric group velocity matching as discussed in~\cite{Bennett2000b,Patera2017}, a configuration that is similar to that adopted for quantum pulse gates~\cite{Brecht2011,Eckstein2011}. In the case where the pump and signal group velocities are matched ($k'_{\mathrm{p}}=k'_{\mathrm{s}}$), the spectral filtering of the phase matching does not limit the time lens aperture that is solely determined by the pump duration. However the temporal walk-off between the idler and the pump introduces a spectral filtering that would be challenging to make negligible. As in the case of symmetric group velocity matching, one could, for example use shorter crystals;
    \item[(iii)] Finite phase-matching and finite aperture: this is the most general case where no restrictions are required on the temporal walk-off and the pump duration. As discussed in~\cite{Bennett2000b,Patera2017} we consider the asymmetric group velocity dispersion matching $k'_{\mathrm{p}}=k'_{\mathrm{s}}$.
\end{enumerate}
\subsubsection*{SVD in the Gaussian kernel approximation}\label{gauss SVD}
In the configurations discussed above, the integral kernel $K(\Omega,\Omega')$
can be approximated by a double Gaussian. This allows to obtain an analytic form for the singular values and the corresponding eigenvectors.

When the group velocity of the pump is matched either to that of the signal or to that of the idler wave (asymmetric group velocity matching, the phase-matching function can be characterized by two bandwidths $\Delta_h$
(along the horizontal direction) and $\Delta_v$ (along the vertical direction) so that \eqref{K}
can be written as
\begin{align}
K(\Omega,\Omega')&\approx
A_{\mathrm{p}}
\mathrm{e}^{-\frac{1}{2}(\Omega-\Omega')^2/\Delta_{\mathrm{p}}^2}
\mathrm{e}^{\frac{\mathrm{i}}{2}D_{\mathrm{f}}(\Omega-\Omega')^2}
\times
\nonumber\\
&\times
\mathrm{e}^{-\frac{1}{2}{\Omega'}^2/\Delta_{\mathrm{h}}^2}
\mathrm{e}^{-\frac{1}{2}{\Omega}^2/\Delta_{v}^2}
\label{K1}
\end{align}
where the two characteristic bandwidths are the inverse of the temporal walk-off between the pump and signal waves ($\Delta_{\mathrm{h}}^{-1}=l_c|k'_{\mathrm{p}}-k'_{\mathrm{i}}|$), or between the pump and the idler waves ($\Delta_{\mathrm{v}}^{-1}=l_c|k'_{\mathrm{p}}-k'_{\mathrm{s}}|$).
\begin{figure*}[t!]
\centering
\subfigure[\label{fig:eval}]{\resizebox{0.6\textwidth}{!}{\includegraphics{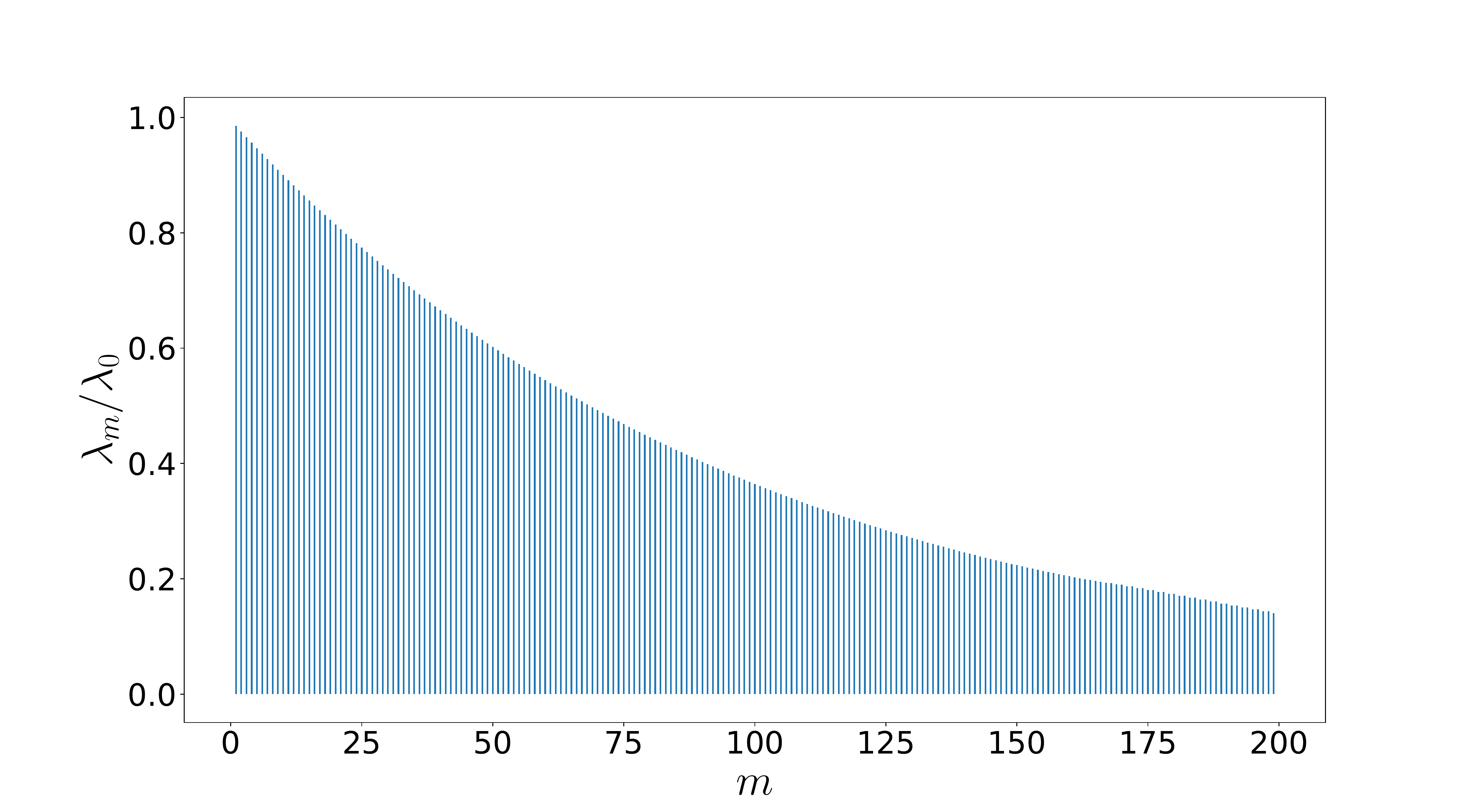}}}\\
\subfigure[\label{fig:evec0out}]{\resizebox{0.4\textwidth}{!}{\includegraphics{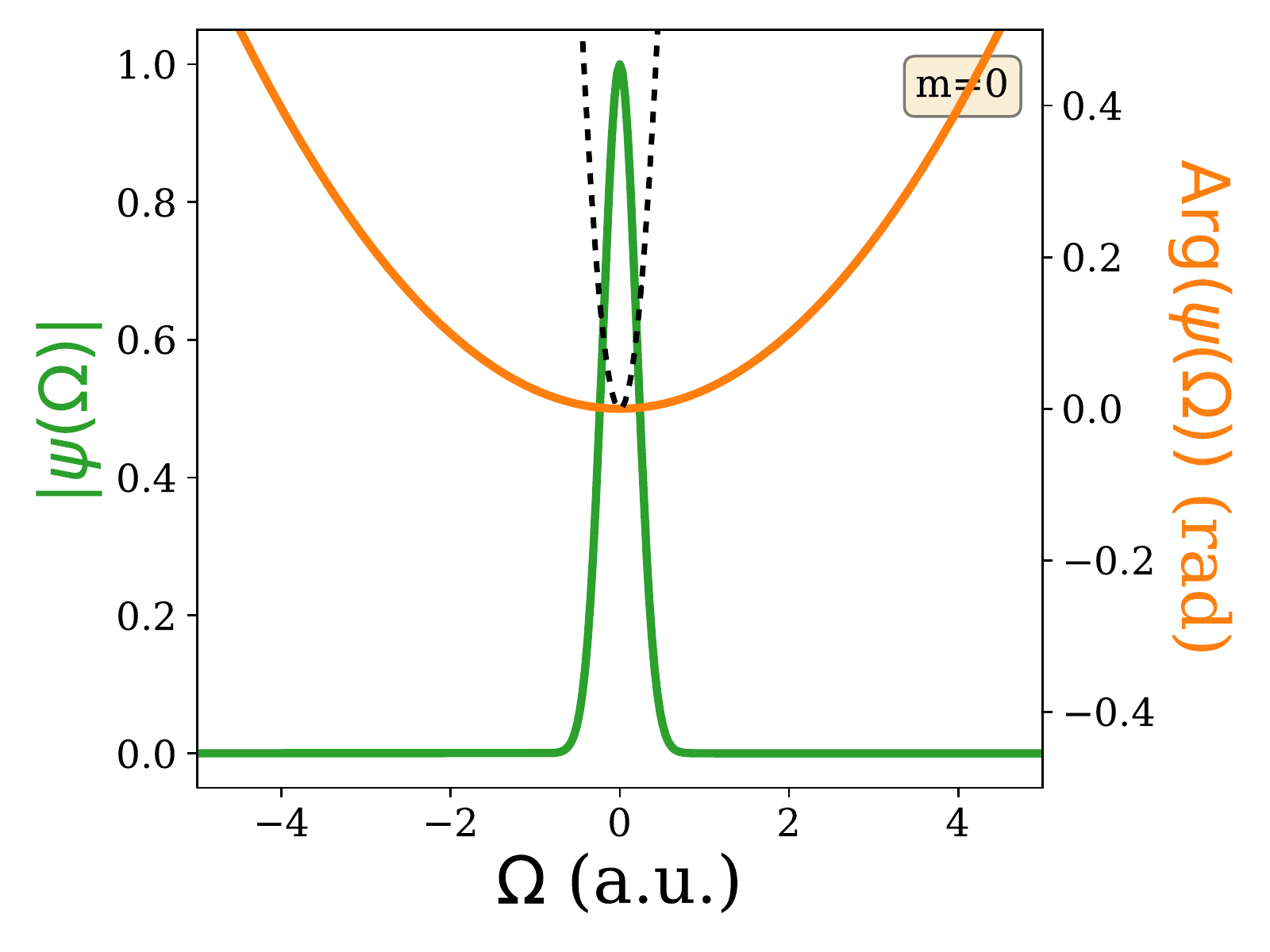}}}
\subfigure[\label{fig:evec1out}]{\resizebox{0.4\textwidth}{!}{\includegraphics{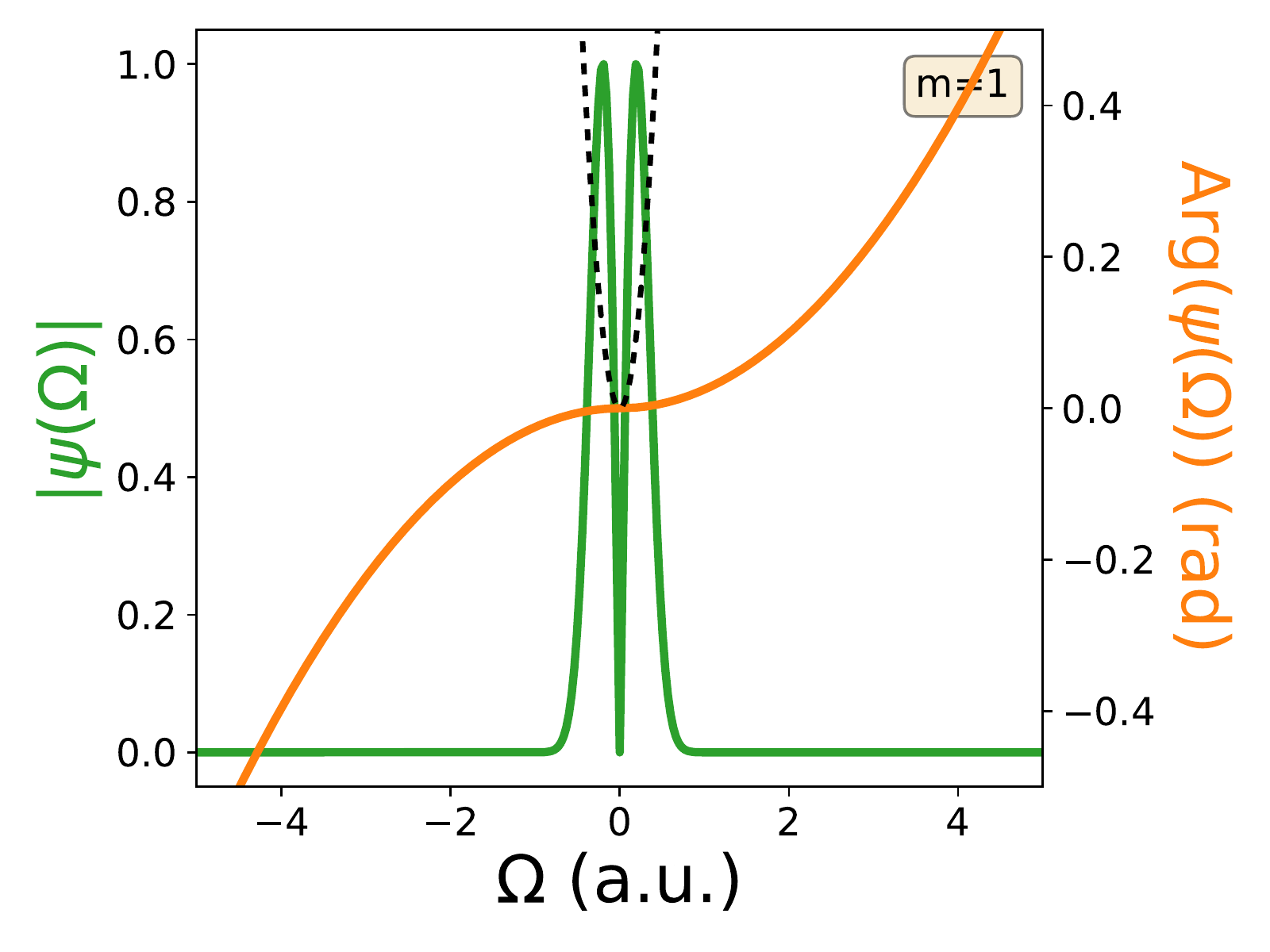}}}\\
\subfigure[\label{fig:evec0in}]{\resizebox{0.4\textwidth}{!}{\includegraphics{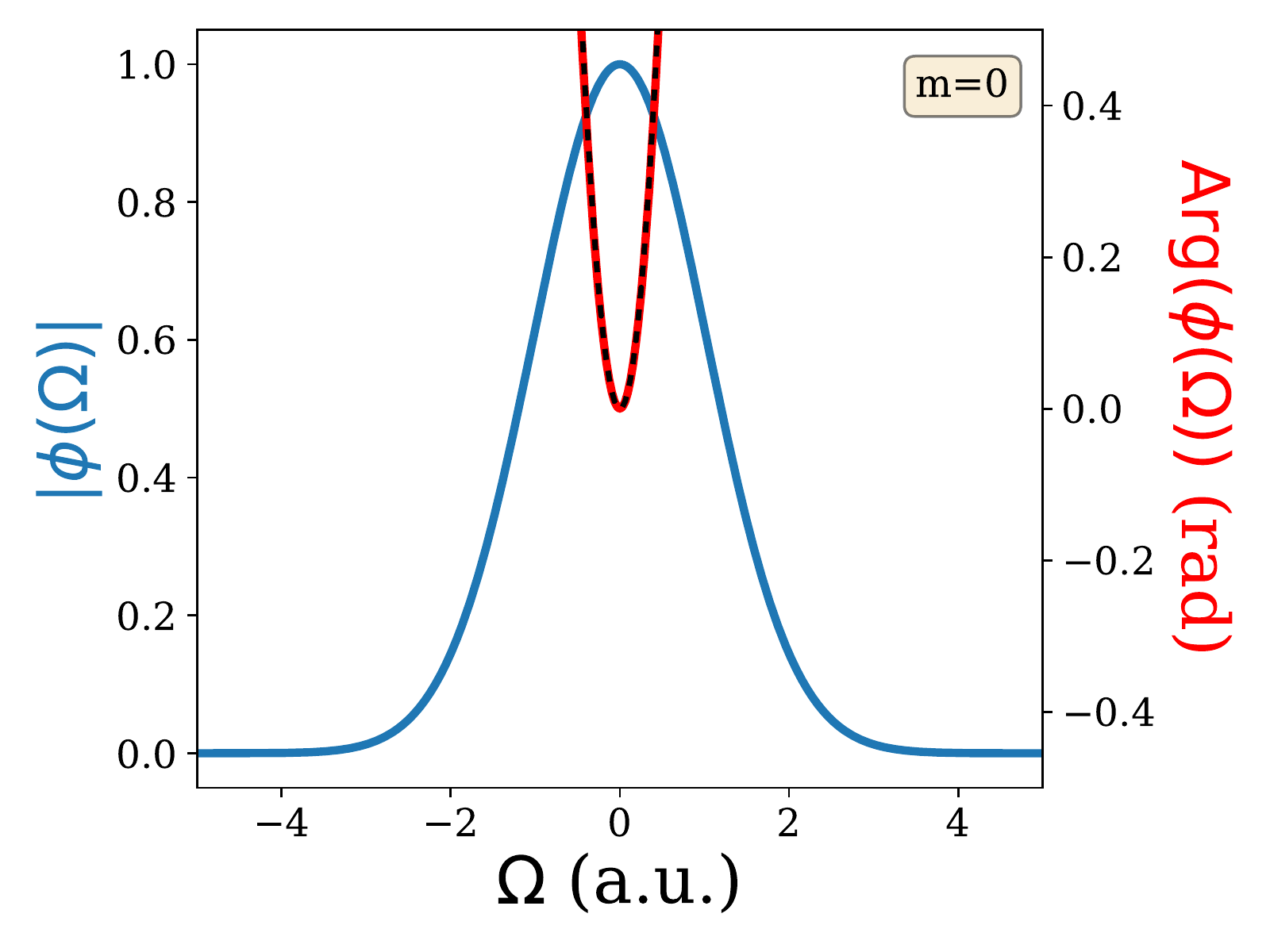}}}
\subfigure[\label{fig:evec1in}]{\resizebox{0.4\textwidth}{!}{\includegraphics{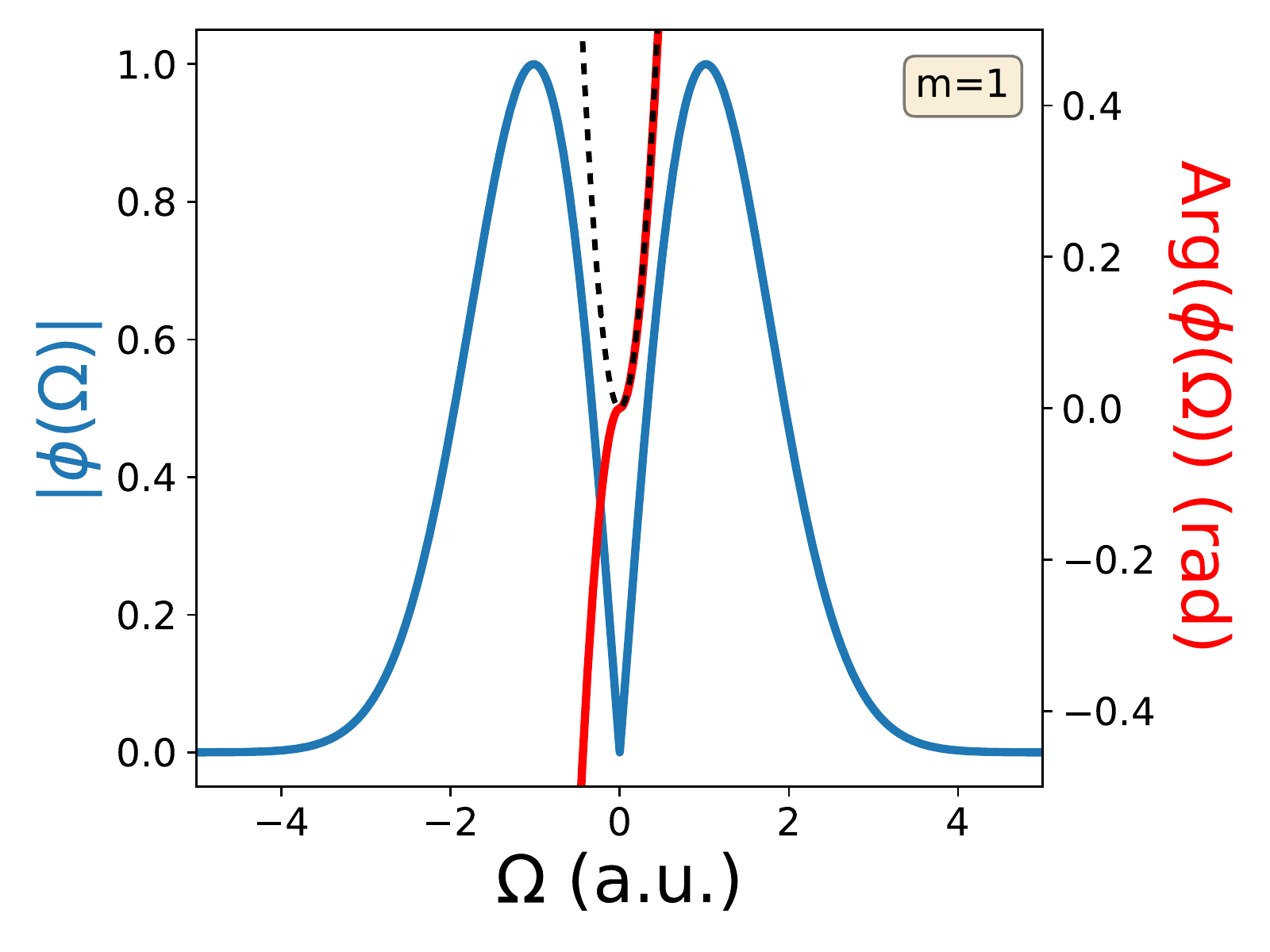}}}
\caption{For parameter choice $\Delta_h=100$ a.u., $\Delta_v=2$ a.u., $\Delta_{\mathrm{p}}=10$ a.u., $D_{\mathrm{f}}=5$ a.u.:
(a) singular values of $K(\Omega,\Omega')$, (b)-(c) module and argument of first two output eigenvectors $m=0$ and $m=1$; (d)-(e) module and argument of first two input eigenvectors $m=0$ and $m=1$. For comparison we trace the quadratic phase $D_{\mathrm{f}}\Omega^2$ induced by the time lens (dashed-black lines).
}
\label{fig:K1 eig}
\end{figure*}
In this case the SVD of $K(\Omega,\Omega')$ leads to the following
analytic results
\begin{align}
\lambda_m&=
A_{\mathrm{p}}
\left[\frac{\pi^{3/2}\widetilde{\Delta}_{h}\widetilde{\Delta}_{v}}{F_1+F_2}
\left(\frac{F_1-F_2}{F_1+F_2}\right)^m\right]^{1/2},\label{eval}
\\
\psi_m(\Omega)&=
\frac{1}{\sqrt{\mathcal{N}_{\psi}}}
\mathrm{e}^{-\frac{\mathrm{i}}{2}D_{\mathrm{f}}(1-\widetilde{\Delta}_{v}^2/\Delta_{\mathrm{p}}^2)\Omega^2}
\mathrm{e}^{-\frac{1}{2}\sigma_{\mathrm{i}}^2\Omega^2}
H_m(\sigma_{\mathrm{i}}\Omega),\label{psi_m}
\\
\phi_m(\Omega)&=
\frac{1}{\sqrt{\mathcal{N}_{\phi}}}
\mathrm{e}^{\frac{\mathrm{i}}{2}D_{\mathrm{f}}(1-\widetilde{\Delta}_{h}^2/\Delta_{\mathrm{p}}^2)\Omega^2}
\mathrm{e}^{-\frac{1}{2}\sigma_{\mathrm{s}}^2\Omega^2}
H_m(\sigma_{\mathrm{s}}\Omega),\label{phi_m}
\end{align}
where $\mathcal{N}_{\phi}$, $\mathcal{N}_{\psi}$ are normalization factors for the eigenmodes,
\begin{align}
F_1&=\sqrt{1+D_{\mathrm{f}}^2 \widetilde{\Delta}_{h}^2 \widetilde{\Delta}_{v}^2},
\label{F1}
\\
F_2&=\sqrt{1-\widetilde{\Delta}_{h}^2 \widetilde{\Delta}_{v}^2/\Delta_{\mathrm{p}}^4},
\label{F2}
\\
\sigma_{\mathrm{i}}&=\sqrt{F_1 F_2}/\widetilde{\Delta}_{v},
\\
\sigma_{\mathrm{s}}&=\sqrt{F_1 F_2}/\widetilde{\Delta}_{\mathrm{h}},
\end{align}
with $1/\widetilde{\Delta}_h^2=1/\Delta_h^2+1/\Delta_{\mathrm{p}}^2$ and $1/\widetilde{\Delta}_v^2=1/\Delta_v^2+1/\Delta_{\mathrm{p}}^2$.

In the typical situation where $\Delta_{v}\ll\Delta_{\mathrm{p}}\ll\Delta_h$, the integral kernel $K(\Omega,\Omega')$
looks like the one depicted in Figure \ref{fig:K1} and $\widetilde{\Delta}_h\approx\Delta_{\mathrm{p}}$ and 
$\widetilde{\Delta}_v\approx\Delta_{v}$. In figures \ref{fig:eval}-\ref{fig:evec1in}, we trace the first 200 singular values, the first two input (blue) and output (green) eigenvectors for the parameters $\Delta_h=100$a.u., $\Delta_v=2$a.u.,  $\Delta_p=10$a.u. and $D_{\mathrm{f}}=5$a.u.

As a final remark we observe that the multimode character of the time lens is grounded in the spectral correlations that come with the chirped pump phase profile (see Figure~\ref{fig:K1 phase}) and make $K(\Omega,\Omega')$ not separable. On the contrary, when the pump is not chirped ($D_{\mathrm{f}}=0$), $K(\Omega,\Omega')$ is separable and the process is single-mode as in the case of quantum pulse gate~\cite{Eckstein2011,Brecht2011}.
\section{Quantum temporal imaging}\label{QTI}
\subsection{Standard approach in the perfect phase-matching approximation}\label{standard QTI}
Under the approximation of perfect phase-matching we get the results of \cite{Patera2018} that we review
in this section for comparison. Since in this case
\begin{align}
f(\Omega,\Omega',\xi)\approx \alpha_{\mathrm{p}}(\Omega-\Omega'),
\end{align}
the right hand side of eqs. \eqref{evo as} and \eqref{evo ai} become convolutions and eq. \eqref{sols} reads
\begin{align}
\left(
\begin{array}{c}
\hat{a}_{\mathrm{s}}(\Omega,l_c/2)
\\
\hat{a}_{\mathrm{i}}(\Omega,l_c/2)
\end{array}
\right)
&=
\mathrm{e}^{M_1(\Omega)\otimes}
\left(
\begin{array}{c}
\hat{a}_{\mathrm{s}}(\Omega,-l_c/2)
\\
\hat{a}_{\mathrm{i}}(\Omega,-l_c/2)
\end{array}
\right)
\label{sols perfectPM}
\end{align}
where $\otimes$ denotes a convolution product.
Differently form the general case, solution \eqref{sols perfectPM} is exact since all the terms of the Magnus expansion higher than the first order are all null.
Then, by using inverse Fourier transform, the transformation \eqref{sols perfectPM} becomes a standard
matrix multiplication
\begin{align}
\left(
\begin{array}{c}
\hat{a}_{\mathrm{s}}(\tau,l_c/2)
\\
\hat{a}_{\mathrm{i}}(\tau,l_c/2)
\end{array}
\right)
&=
B(\tau)
\left(
\begin{array}{c}
\hat{a}_{\mathrm{s}}(\tau,-l_c/2)
\\
\hat{a}_{\mathrm{i}}(\tau,-l_c/2)
\end{array}
\right)
\label{sols II perfectPM}
\end{align}
where
\begin{align}
B(\tau)
&=
\left(
\begin{array}{cc}
c(\tau) & \mathrm{e}^{-\mathrm{i}\phi(\tau)}s(\tau)
\\
-\mathrm{e}^{\mathrm{i}\phi(\tau)} s(\tau) & c(\tau)
\end{array}
\right)
\label{Bt}
\end{align}
with $\phi(\tau)=\mathrm{Arg}[\tilde{\alpha}_{\mathrm{p}}(\tau)]$, $\tilde{\alpha}_{\mathrm{p}}(\tau)$ the inverse Fourier transform of $\tilde{\alpha}_{\mathrm{p}}(\Omega)$
\begin{align}
\tilde{\alpha}_{\mathrm{p}}(\tau)&\propto
A_{\mathrm{p}}\,
\mathrm{e}^{-\tau^2/2{\tau'_{\mathrm{p}}}^2}
\mathrm{e}^{\mathrm{i}\tau^2/2D_{\mathrm{f}}}
\end{align}

and
\begin{align}
c(\tau)&=\cos(g l_c |\tilde{\alpha}_{\mathrm{p}}(\tau)|),
\\
s(\tau)&=\sin(g l_c |\tilde{\alpha}_{\mathrm{p}}(\tau)|).
\end{align}
In particular the idler wave at the output ($z=l_c/2$) of the SFG process is given by
\begin{align}
\hat{a}_{\mathrm{i}}(\tau,l_c/2)=
&-\mathrm{e}^{\mathrm{i}\phi(\tau)}s(\tau)\hat{a}_{\mathrm{s}}(\tau,-l_c/2)
+
c(\tau)\hat{a}_{\mathrm{i}}(\tau,-l_c/2),
\label{idler out SFG perfectPM}
\end{align}
Equation \eqref{sols II perfectPM} represents a unitary transformation of the photon annihilation operators from the input of the nonlinear crystal to
its output, hence preserving the canonical commutation relations. This equation has the same form as the transformation induced by a beam splitter
with the amplitude transmission coefficient $c(\tau)$ and the reflection coefficient $s(\tau)$
such that $c(\tau)^2+s(\tau)^2=1$. The transmission coefficient controls the amount of the input waves that remain in the same mode and the reflection coefficient controls the amount the input waves that is converted in the other mode.
The phase factor $\mathrm{e}^{\mathrm{i}\phi(\tau)}$ in front of the input signal amplitude is
determined by the phase of the pump wave. Therefore a time lens transformation is obtained by choosing a quadratic time dependence in the pump phase
$\phi(\tau)=\tau^2/2D_{\mathrm{f}}$. Notice that the second term at the right-hand side of \eqref{idler out SFG perfectPM} is associated to
vacuum fluctuations entering the nonlinear process through the input idler port of the time lens and mixing with the input state. 
These fluctuations are, of course, detrimental for the nonclassical input states and they need to be avoided. They can be suppressed
when the conversion efficiency of the process $|s(\tau)|^2=1$. This situation can be reached when $g l_c |\alpha_{\mathrm{p}}(\tau)|=\pi/2$. However this condition
cannot be satisfied for all $\tau$ because of a finite duration of the pump pulse. Typically the pump pulse presents a maximum intensity at $\tau=0$,
then the conversion efficiency of the nonlinear process can be optimized such as 
\begin{align}
g l_c |\alpha_{\mathrm{p}}(0)|=\pi/2,
\label{pi2}
\end{align}
while for $\tau\neq0$ the conversion efficiency will be smaller than one: as a consequence the time lens presents a finite aperture of the imaging scheme.

The linear unitary transformation for the imaging scheme considered in figure \ref{fig:imag scheme} can be obtained by applying, one
after the other, the transformation \eqref{disp evo} for the input dispersive propagation with GDD $D_\mathrm{in}$, 
then \eqref{idler out SFG perfectPM} for the time lens and finally \eqref{disp evo} for the output dispersive propagation with GDD $D_\mathrm{out}$:
\begin{align}
\hat{A}_{\mathrm{i}}^{\mathrm{out}}(\tau)&=
\int\mathrm{d}\tau'\,\left[
h_{\mathrm{i}}(\tau,\tau')\hat{A}_{\mathrm{s}}^{\mathrm{in}}(\tau')
+
q_{\mathrm{i}}(\tau,\tau')\hat{A}_{\mathrm{i}}^{\mathrm{in}}(\tau')
\right]
\label{Ai out perfectPM}
\end{align}
where
\begin{align}
h_{\mathrm{i}}(\tau,\tau')&=
-\int\mathrm{d}\tau''\,
G_{\mathrm{out}}(\tau-\tau'')
s(\tau'')\mathrm{e}^{\mathrm{i}\tau''^2/2D_{\mathrm{f}}}
\times
\nonumber\\
&\quad\quad\quad\quad\times
G_{\mathrm{in}}(\tau''-\tau'),
\\
q_{\mathrm{i}}(\tau,\tau')&=G_{\mathrm{out}}(\tau-\tau')c(\tau'),
\end{align}
are the IRFs of the transformation.
In \cite{Patera2018} we showed that under the Goodman-Tichenor approximation and when the \textit{imaging condition}
\begin{align}
\frac{1}{D_{\mathrm{in}}}+\frac{1}{D_{\mathrm{out}}}=\frac{1}{D_{\mathrm{f}}}
\label{imag cond}
\end{align}
is satisfied, then the IRFs become
\begin{align}
 h_{\mathrm{i}}(\tau,\tau')&=
-\mathrm{i}\sqrt{|M|}\mathrm{e}^{\frac{-\mathrm{i}\tau^2}{2|M|D_{\mathrm{f}}}}
\int\frac{\mathrm{d}\Omega}{2\pi}\mathrm{e}^{\mathrm{i}\Omega(\tau-M\tau')}s(D_{\mathrm{out}}\Omega),
\\
q_{\mathrm{i}}(\tau,\tau')&=
\mathrm{i}\sqrt{|M|}\mathrm{e}^{\frac{-\mathrm{i}\tau^2}{2|M|D_{\mathrm{f}}}}
\int\frac{\mathrm{d}\Omega}{2\pi}\mathrm{e}^{\mathrm{i}\Omega(\tau-M\tau')}\tilde{c}(D_{\mathrm{out}}\Omega)
\end{align}
where $M=-D_{\mathrm{out}}/D_{\mathrm{in}}$ is the magnification factor of the imaging scheme and 
$\tilde{c}(\tau)=c(\tau)\mathrm{exp}(-\mathrm{i}\tau^2/2D_{\mathrm{f}})$.

Notice that the response $h_i(\tau,\tau')$ corresponds to the classical the IRF and it is given, as in standard imaging, by the Fourier transform of the pupil function that in our
case is the function $s(\tau)$. On the other side the response $q_i(\tau,\tau')$ has no classical correspondence
and it is responsible for the vacuum fluctuations entering the scheme because of a finite pupil function.

When the pump pulse is infinitely long, hence the temporal aperture of the lens is arbitrarily large, and the condition \eqref{pi2} is satisfied, then
$s(\tau)=1$ and $c(\tau)=0$. In this case the response functions become ideal
\begin{align}
 h_{\mathrm{i}}(\tau,\tau')&=
-\mathrm{i}\sqrt{|M|}\mathrm{e}^{\frac{-\mathrm{i}\tau^2}{2|M|D_{\mathrm{f}}}}
\delta(\tau-M\tau'),
\\
q_{\mathrm{i}}(\tau,\tau')&=0
\end{align}
such that the transformation \eqref{Ai out perfectPM}
\begin{align}
\hat{A}_{\mathrm{i}}^{\mathrm{out}}(\tau)&=
-\mathrm{i}\frac{\mathrm{e}^{\frac{-\mathrm{i}\tau^2}{2|M|D_{\mathrm{f}}}}}{\sqrt{|M|}}
\hat{A}_{\mathrm{i}}^{\mathrm{in}}(\tau/M)
\end{align}
describes an ideal imaging scheme as that considered in \cite{Zhu2013,Patera2015}.

\subsection{Modal approach}
For a non ideal phase-matching (see configuration (iii) in section II) it is not possible to analitically solve the propagation of field amplitudes through the time lens 
(eqs. \eqref{evo as} and \eqref{evo ai}) and a perturbative approach is required. In section \ref{Timelens} we used the first order of the
Magnus expansion that is suitable for high conversion efficency regimes.
This analisys leads to solutions \eqref{sols II} that are expressed in terms of the singular values and eigenfunctions of the kernel
$K(\Omega,\Omega')$ (see eq. \eqref{Kbis}). This explains the necessity of a modal approach to quantum temporal imaging.

The imaging transformation is obtained by applying one
after the other the transformations eq. \eqref{disp evo} for the input dispersive propagation,
eq. \eqref{sols II} for the time lens and eq. \eqref{disp evo} for the output dispersive propagation:
\begin{align}
\hat{A}_{\mathrm{i}}^{\mathrm{out}}(\tau)
&=
\int\frac{\mathrm{d}\Omega}{\sqrt{2\pi}}\,
\mathrm{e}^{-\mathrm{i}\Omega\tau}
\hat{A}_{\mathrm{i}}^{\mathrm{out}}(\Omega)
\label{Ai out}
\\
\hat{A}_{\mathrm{i}}^ {\mathrm{out}}(\Omega)
&=
\int\mathrm{d}\Omega'\,
\Big[
h_{\mathrm{i}}(\Omega,\Omega')\hat{A}_{\mathrm{s}}^{\mathrm{in}}(\Omega')
+
q_{\mathrm{i}}(\Omega,\Omega')\hat{A}_{\mathrm{i}}^{\mathrm{in}}(\Omega')
\Big]
\label{Ai out FT}
\end{align}
where $h_{\mathrm{i}}(\Omega,\Omega')$ and $q_{\mathrm{i}}(\Omega,\Omega')$ are the two transfer functions given by
\begin{align}
h_{\mathrm{i}}(\Omega,\Omega')&=
-\mathcal{G}_{\mathrm{out}}(\Omega)V_{\mathrm{i}}(\Omega,\Omega')\mathcal{G}_{\mathrm{in}}(\Omega'),
\label{hi}
\\
q_{\mathrm{i}}(\Omega,\Omega')&=
\mathcal{G}_{\mathrm{out}}(\Omega)U_{\mathrm{i}}(\Omega,\Omega').
\label{qi}
\end{align}
The classical IRF that allows to quantify the system performances is then obtained by a Fourier transform of expression~\eqref{hi}. We note that this is not a difficult task in the Gaussian kernel approximation since we deal with the Fourier transform of Gauss-Hermite functions.

Since the functions \eqref{hi} and \eqref{qi} satisfy the relation
\begin{align}\label{unit cond}
\int\mathrm{d}\Omega''\,
\Big[
h_{\mathrm{i}}(\Omega,\Omega'')h_{\mathrm{i}}^*(\Omega',\Omega'')
&+
\,q_{\mathrm{i}}(\Omega,\Omega'')q_{\mathrm{i}}^*(\Omega',\Omega'')
\Big]
=
\nonumber\\
&=\delta(\Omega-\Omega')
\end{align}
then \eqref{Ai out} is unitary and the field commutators are preserved at the output of the scheme
\begin{align}
\left[
\hat{A}_{\mathrm{i}}^{\mathrm{out}}(\tau),\left.\hat{A}_{\mathrm{i}}^{\mathrm{out}}\right.^{\dag}(\tau')
\right]
&=
\delta(\tau-\tau').
\end{align}
By using expressions \eqref{Ui} and \eqref{Vi}, it is possible to write \eqref{hi} and \eqref{qi}
in diagonal form
\begin{align}
h_{\mathrm{i}}(\Omega,\Omega')&=
\sum_m (-s_m) \xi_m(\Omega)\zeta_m^*(\Omega'),
\label{hi diag}
\\
q_{\mathrm{i}}(\Omega,\Omega')&=
\sum_m c_m \xi_m(\Omega)\psi_m^*(\Omega'),
\label{qi diag}
\end{align}
where $s_m=\sin(g l_c \lambda_m)$, $c_m=\cos(g l_c \lambda_m)$ and
\begin{align}
\xi_m(\Omega)&=\mathcal{G}_{\mathrm{out}}(\Omega)\psi_m(\Omega),
\\
\zeta_m(\Omega)&=\mathcal{G}_{\mathrm{in}}^ *(\Omega)\phi_m(\Omega).
\end{align}
Notice that the family of functions $\{\xi_m\}$ and $\{\zeta_m\}$ still form two complete sets of orthonormal functions. 

Even if it is not interesting from the point of view of quantum temporal imaging, for completeness we consider below also 
the unitary transformation describing the output signal mode. This results to be
\begin{align}
\hat{A}_{\mathrm{s}}^{\mathrm{out}}(\tau)
&=
\int\frac{\mathrm{d}\Omega}{\sqrt{2\pi}}\,
\mathrm{e}^{-\mathrm{i}\Omega\tau}
\hat{A}_{\mathrm{s}}^{\mathrm{out}}(\Omega)
\label{As out}
\\
\hat{A}_{\mathrm{s}}^ {\mathrm{out}}(\Omega)
&=
\int\mathrm{d}\Omega'\,
\Big[
h_{\mathrm{s}}(\Omega,\Omega')\hat{A}_{\mathrm{i}}^{\mathrm{in}}(\Omega')
+
q_{\mathrm{s}}(\Omega,\Omega')\hat{A}_{\mathrm{s}}^{\mathrm{in}}(\Omega')
\Big]
\label{As out FT}
\end{align}
where, after using expressions \eqref{Us} and \eqref{Vs},
\begin{align}
h_{\mathrm{s}}(\Omega,\Omega')&=
\sum_m s_m \phi_m(\Omega)\psi_m^*(\Omega'),
\label{hs diag}
\\
q_{\mathrm{s}}(\Omega,\Omega')&=
\sum_m c_m \phi_m(\Omega)\zeta_m^*(\Omega').
\label{qs diag}
\end{align}
The ensemble of expressions \eqref{hi diag}, \eqref{qi diag}, \eqref{hs diag} and \eqref{qs diag}  
represent the decomposition of the full transformation associated to the imaging scheme in terms of singular values and eigenvectors.

Because of the completeness  of $\{\zeta_m\}$ an input quantum image can be decomposed as
\begin{align}
\hat{A}_{\mathrm{s}}^{\mathrm{in}}(\Omega)&=
\sum_m \hat{A}_{\mathrm{s},m}^\mathrm{in}\zeta_m(\Omega),
\end{align}
hence the functions $\zeta_m$ can be regarded as degrees of freedom of the input object.
Also the input vacuum field can be decomposed as
\begin{align}
\hat{A}_{\mathrm{i}}^{\mathrm{in}}(\Omega)&=
\sum_m \hat{A}_{\mathrm{i},m}^\mathrm{in}\psi_m(\Omega).
\end{align}
Then, because of completeness of $\{\xi_m\}$, we can write the expansion
\begin{align}
\hat{A}_{\mathrm{i}}^{\mathrm{out}}(\Omega)&=
\left(
\sum_m 
\left(-s_m \hat{A}_{\mathrm{s},m}^\mathrm{in}
+
c_m \hat{A}_{\mathrm{i},m}^\mathrm{in}\right)
\right)
\xi_m(\Omega).
\label{Aout decomp}
\end{align}
Typical experimental implementations, like that discussed in \ref{gauss SVD}, 
are far from the ideal situation since the signal-to-idler conversion efficiencies $s_m$ fall off for increasing values of $m$.
Therefore, in a general situation, the higher modal components of the signal object will be mixed with vacuum fluctuations 
at the output of the imaging scheme.
Since the singular values \eqref{eval} of the process depend on the pump amplitude $A_{\mathrm{p}}$, it is possible to choose this value such that
a particular $s_m$ is equal to one, but this condition cannot be satisfied for all the
rest of the eigenspectrum. In the following we will consider a pump amplitude such that the fundamental modal component $m=0$ presents
$100\%$ conversion efficiency in the 1st order Magnus perturbation theory ($s_0=1$ and $c_0=0$); this condition is reached when
\begin{align}\label{pi2bis}
g l_c \lambda_0 = \pi/2,
\end{align}
being $\lambda_0$ the most important singular value.
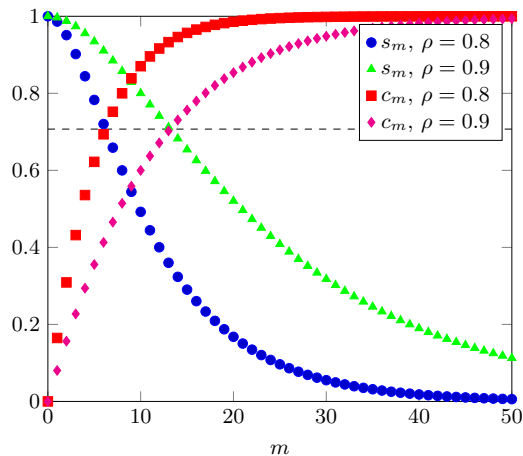
\begin{figure}[t!]
\begin{center}
\begin{tikzpicture}[scale=0.9]
\begin{axis}
[
xlabel={$m$},
ylabel={},
xmin=0, xmax=50,
ymin=0, ymax=1,
]
\addplot+[ycomb, only marks, domain=1:10, samples at={0,1,...,50}, blue]{sin(2*atan(1)*(0.8)^(x/2))};
\addlegendentry{$s_m$, $\rho=0.8$}
\addplot[ycomb, only marks, mark=triangle*, green, domain=1:10, samples at={0,1,...,50}]{sin(2*atan(1)*(0.9)^(x/2))};
\addlegendentry{$s_m$, $\rho=0.9$}
\addplot[ycomb, only marks, mark=square*, red, domain=1:10, samples at={0,1,...,50}]{cos(2*atan(1)*(0.8)^(x/2))};
\addlegendentry{$c_m$, $\rho=0.8$}
\addplot[ycomb, only marks, mark=diamond*, domain=1:10, samples at={0,1,...,50}, magenta]{cos(2*atan(1)*(0.9)^(x/2))};
\addlegendentry{$c_m$, $\rho=0.9$}
\addplot[domain=0:50, dashed, black]{cos(atan(1))};
\end{axis}
\end{tikzpicture}
\end{center}
\caption{Plot of $s_m=\sin(\lambda_m)$ (blue-circles and green-triangles) and $c_m=\cos(\lambda_m)$ (red-squares and magenta-diamonds) for 
$g l_c\lambda_0=\pi/2$ and $\rho=\frac{F_1-F_2}{F_1+F_2}\in\{0.8,0.9\}$.}
\label{smcm}
\end{figure}
\begin{figure*}[t!]
\centering
\subfigure[\label{fig:Vi module}]{\resizebox{0.45\textwidth}{!}{\includegraphics{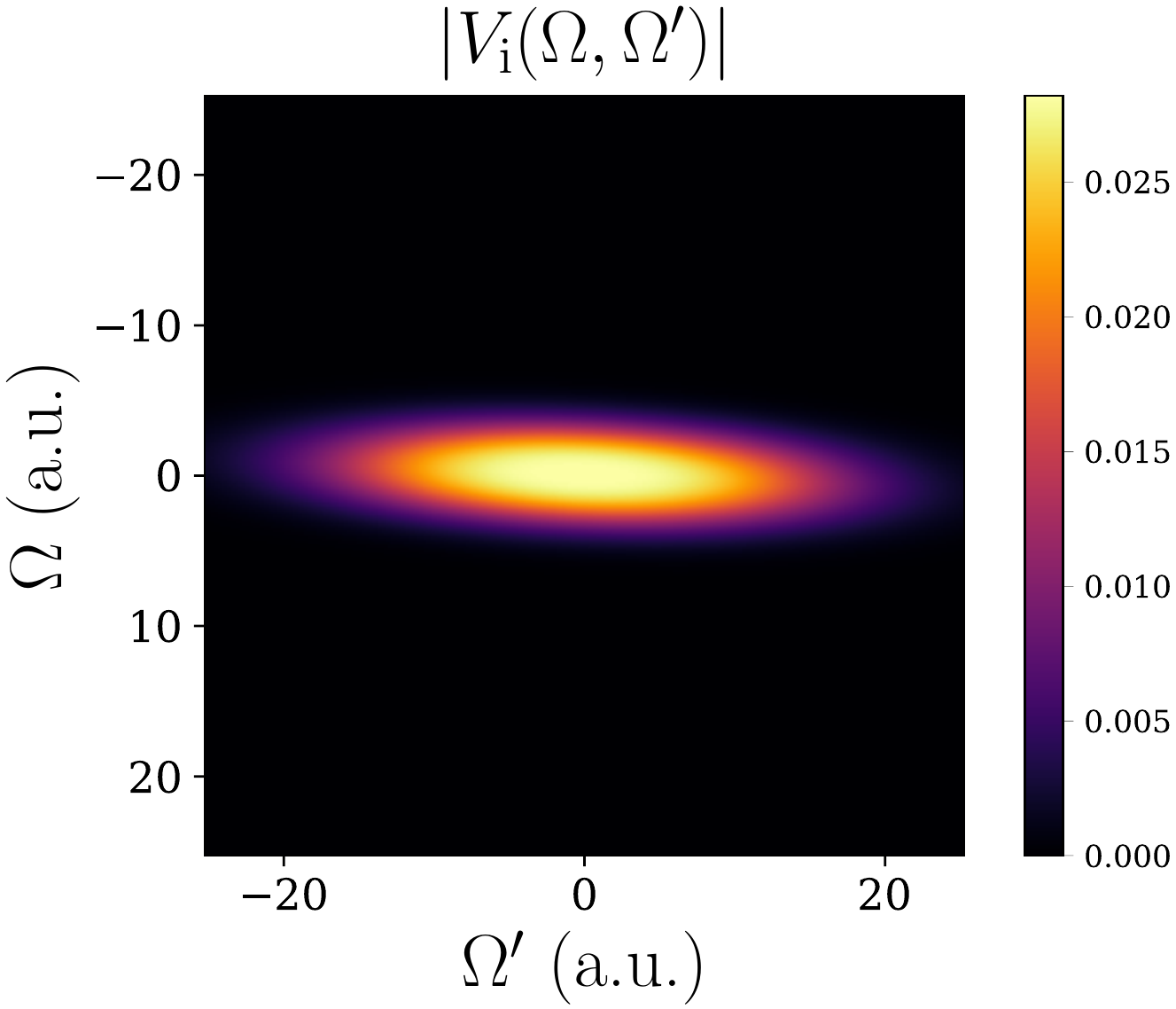}}}
\quad
\subfigure[\label{fig:Vi phase}]{\resizebox{0.45\textwidth}{!}{\includegraphics{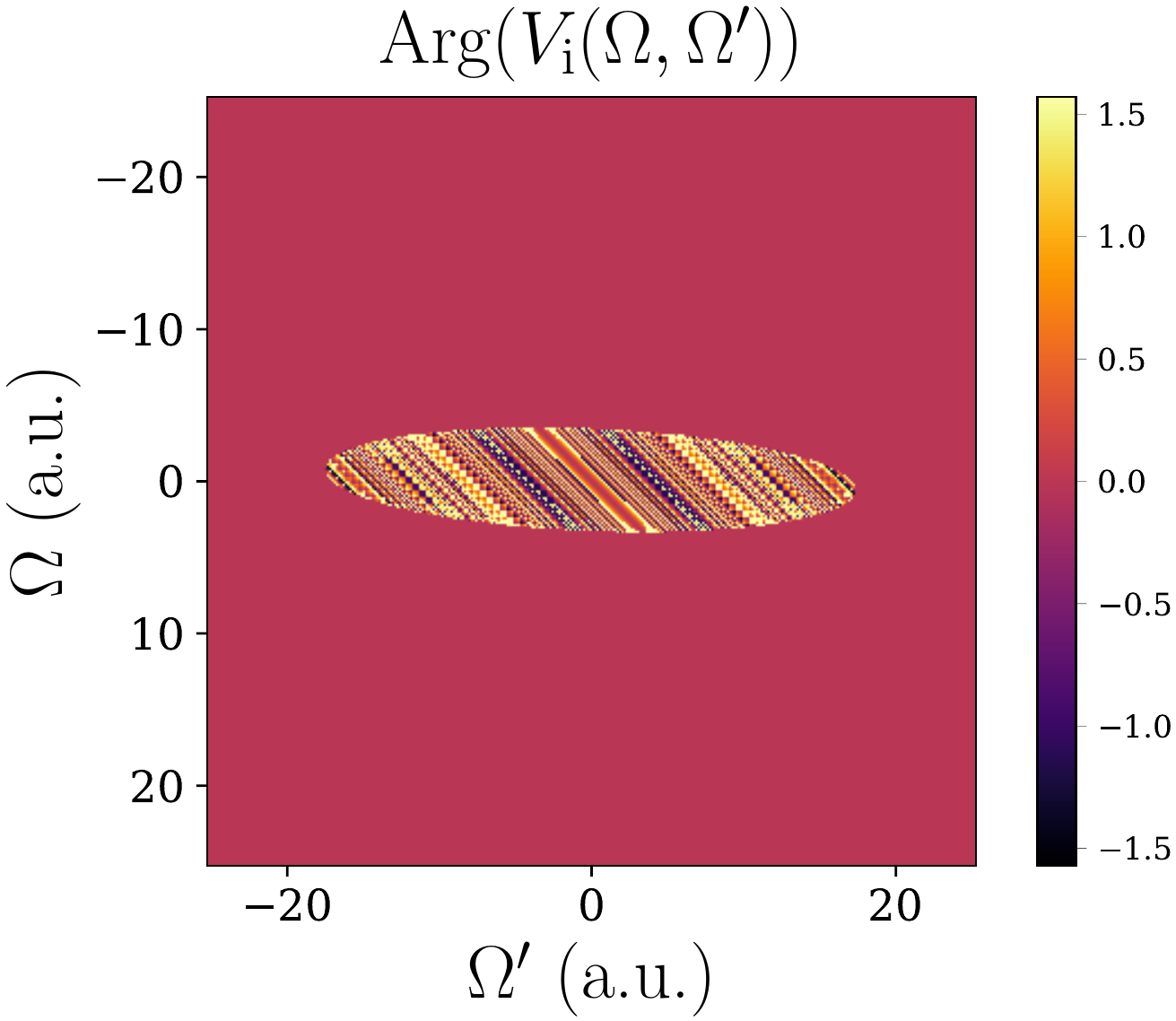}}} \\
\caption{(a) module and (b) phase of $V_{\mathrm{i}}(\Omega,\Omega')$ as described in eq. \eqref{Vi}. 
The parameters are: $\Delta_h=100$ a.u., $\Delta_v=2$ a.u., $\Delta_{\mathrm{p}}=10$ a.u., $D_{\mathrm{f}}=5$ a.u.}
\label{fig:Vi}
\end{figure*}
In figure \ref{smcm} we trace the values of coefficients $s_m$ and $c_m$ for an experimental situation corresponding
to condition \eqref{pi2bis}. From this figure it is clear that while for the first eigenmode
the situation is ideal, it rapidly get worse for all the other modes: the higher the order $m$ of the eigenmode,
the higher the contribution $c_m$ of the input vacuum fluctuations in the idler channel. 

The modal analysis we have performed shows that the properties of the imaging transformation \eqref{Ai out FT} are completely described
by the set of coefficients ${s_m}$ and ${c_m}$ and by the family of eigenmodes $\{\phi_m,\psi_m,\xi_m,\zeta_m\}$. This fact allows to design optimal experimental configurations. Notice indeed that for a setup corresponding to lager values of the ratio
\begin{equation}
\rho=\frac{F_1-F_2}{F_1+F_2}    
\end{equation}
the roll-off of the $s_m$ coefficients is less important so that a larger number of modes is not corrupted by vacuum noise. One can appreciate this difference in figure~\ref{smcm}, where the coefficients $s_m$ for a $\rho=0.9$ (green triangles) are compared to those corresponding to a smaller value $\rho=0.8$ (blue dots).
Since the parameters $F_1$ and $F_2$ depend on experimentally controllable parameters (see Eqs.~\eqref{F1} and~\eqref{F2}), highly multimode setups that are not dominated by quantum noise could be designed.
\section{Quantification of system performances}
\begin{figure*}[t!]
\centering
\subfigure[\label{fig:h sec}]{\resizebox{0.4\textwidth}{!}{\includegraphics{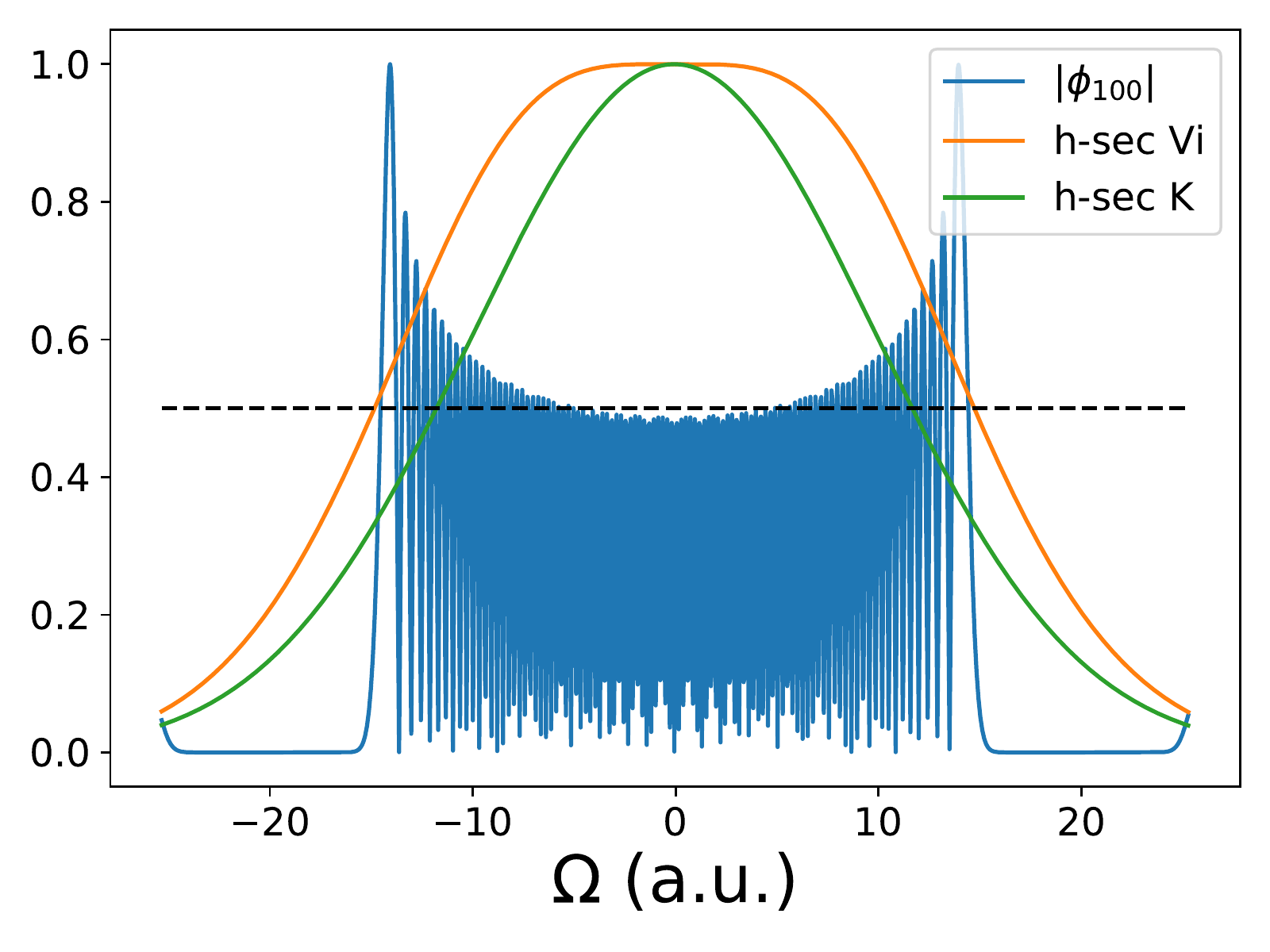}}} \quad
\subfigure[\label{fig:v sec}]{\resizebox{0.4\textwidth}{!}{\includegraphics{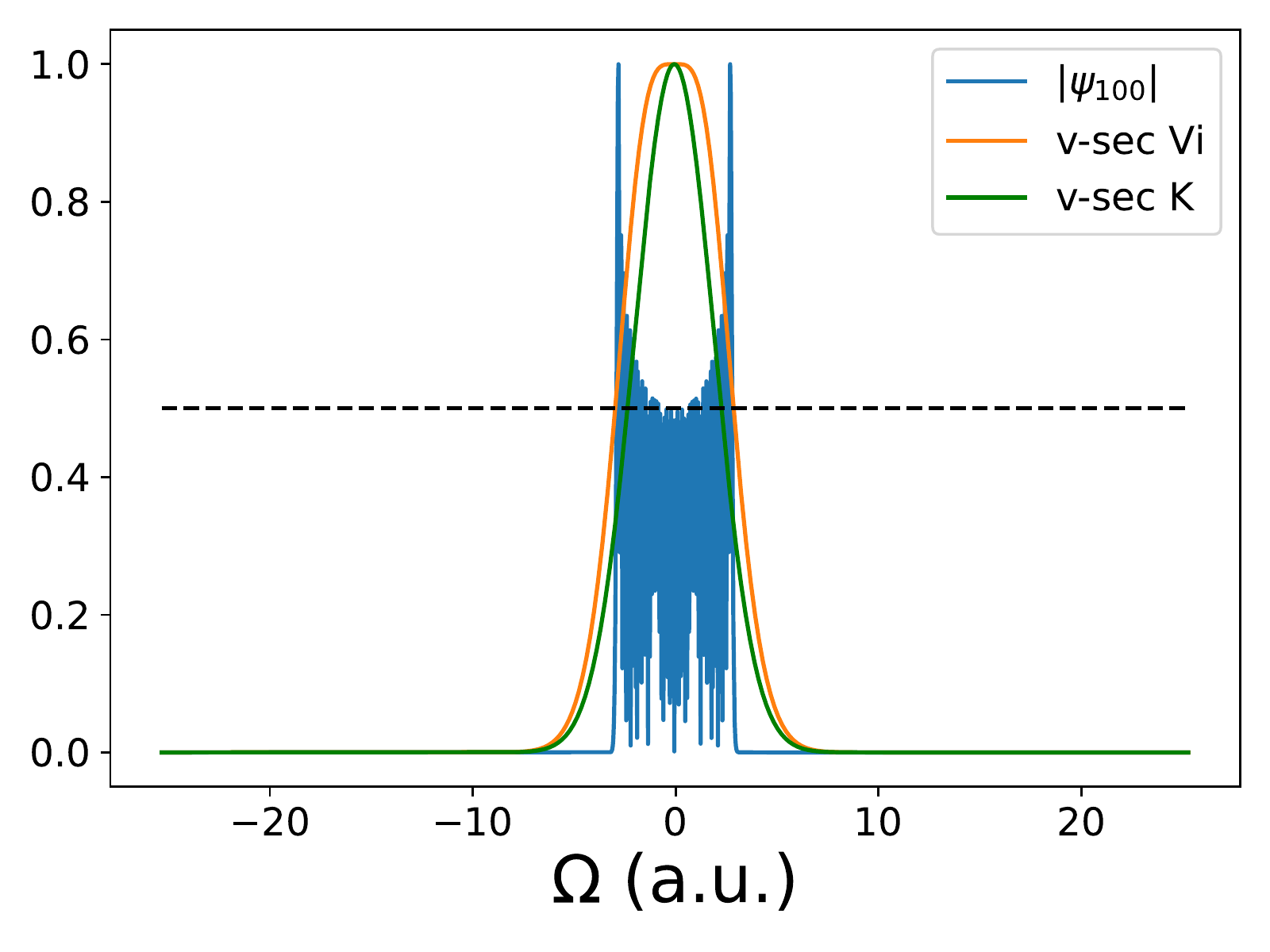}}}
\caption{For parameter choice $\Delta_h=100$ a.u., $\Delta_v=2$ a.u., $\Delta_{\mathrm{p}}=10$ a.u., $D_{\mathrm{f}}=5$ a.u.:
(a) Comparison between the horizontal section ($\Omega=0$) of $K(\Omega,\Omega')$ (green solid) and that of $V_{\mathrm{i}}(\Omega,\Omega')$ (yellow solid);
the full width at half maximum of the horizontal section of $V_{\mathrm{i}}(\Omega,\Omega')$ is close to the FWHM of the eigenmode $\phi_m(\Omega)$ with
$m=S$ (in this case $S=100$);
(b) Comparison between the vertical section ($\Omega'=0$) of $K(\Omega,\Omega')$ (green solid) and that of $V_{\mathrm{i}}(\Omega,\Omega')$ (yellow solid);
the full width at half maximum of the vertical section of $V_{\mathrm{i}}(\Omega,\Omega')$ is close to the FWHM of the absolute value of eigenmode $|\psi_m(\Omega)|$ with $m=S$ (with $S=100$);
}
\label{fig:hvsec}
\end{figure*}
While the performances of a temporal imaging scheme can be easily quantified in the case of perfect phase-matching for any 
conversion efficiency regime, as soon as one wants to include the limitations induced by the finite phase-matching the quantification of
performances becomes a difficult task. Despite the fact that the expressions \eqref{hi} and \eqref{qi} do not allow for an analytic evaluation 
of the transfer functions $h_{\mathrm{i}}(\Omega,\Omega')$, $q_{\mathrm{i}}(\Omega,\Omega')$, we show in this section that
a quantitative assessment of the performances can be realized from the singular values and eigenvectors discussed in the previous
section thus showing the interest of the modal approach for temporal imaging.

As discussed by Bennett and Kolner in \cite{Bennett2000b}, a finite Group Velocity Mismatch (GVM) between the three waves involved 
in the SFG process results in a spectral filtering that limits the bandwidth of the transmitted field amplitude and modifies the ideal
impluse response $\tilde{h}(\tau)$. The resolution is now determined by the width of the effective impulse response $\tilde{h}'(\tau)$ that
can be obtained, in two simplifying cases, as the convolution the ideal impulse response $\tilde{h}(\tau)$
with the inverse Fourier transform of a spectral filtering function $\mathcal{F}(\tau)$: 
when the group velocity of the pump matches that of the signal, the GVM between pump and ildler has the effect equivalent to a filter at the output. 
Hence the total impulse response is
\begin{align}
\tilde{h}'(\tau) &\propto
\mathcal{F}_{\mathrm{out}}(\tau)\otimes\tilde{h}(\tau).
\end{align}
On the other side, when the group velocity of the pump matches that of the idler, the GVM between pump and signal has the effect equivalent to
an input filter such that the total impulse response is
\begin{align}
\tilde{h}'(\tau) &\propto
\tilde{h}(\tau)\otimes \mathcal{F}_{\mathrm{in}}(\tau/M),
\end{align}
with $M=-D_{\mathrm{out}}/D_{\mathrm{in}}$ the magnification factor of the imaging scheme.

In the low conversion efficiency regime, as the one considered by Bennett and Kolner, the filtering functions have an analytic expression, 
therfore the bandwidth of their inverse Fourier transform can be obtained as $|k'_{\mathrm{p}}-k'_{\mathrm{s}}|l_c$ (respectively $|k'_{\mathrm{p}}-k'_{\mathrm{i}}|l_c$).
In the high conversion efficiency regimes the time lens transformation \eqref{Vi} is significantly different from that in the low efficiency regime \eqref{K},
therefore the approach of \cite{Bennett2000b} is less precise. This difference can be appreciated by comparing figures \ref{fig:K1} and \ref{fig:Vi}:  
at the first order of the Magnus expansion the phase matching has no more the profile of a double Gaussian as $K(\Omega,\Omega')$ and the spectral region
where it is maximal is larger and flatter (compare figures \ref{fig:K1 module} and \ref{fig:Vi module}). Also this difference can be observed in figure \ref{fig:hvsec}
where the horizontal ($\Omega=0$) and the vertical ($\Omega'=0$) sections of $K(\Omega,\Omega')$ and $V_{\mathrm{i}}(\Omega,\Omega')$ are compared.
On the other side the phase profile of $V_{\mathrm{i}}(\Omega,\Omega')$
(see figure \ref{fig:Vi phase}) shows that the linear chirp induced by the pump is still present. This fact ensures that the scheme still works as a time lens in the high conversion efficiency regime.

\vspace{10pt}
The starting point of the modal approach consists in estimating the extension of spectrum of the singular values $\{s_m\}$ by means of the Schmidt number $S$ that is defined as
\begin{align}
S=\frac{\left(\sum_m s_m^2\right)^2}{\sum_m s_m^4}
\end{align}
and characterizes the amount of degrees of freedom of the transformation \eqref{Ai out FT}.

\vspace{10pt}
\textit{Spectral Field of View}: by the knowledge of the Schmidt number $S$ we can get also an estimate of the bandwidths of the horizontal ($\Delta_{\mathrm{s}}$) 
and vertical ($\Delta_{\mathrm{s}}$) sections of $|V_{\mathrm{i}}(\Omega,\Omega')|$. 
Notice that these bandwidths characterize also the transfer function $h_{\mathrm{i}}(\Omega,\Omega')$ by virtue of expression $\eqref{hi}$. 
These bandwidths can be approximately obtained as the variances of the eigenfunctions $\zeta_{m}(\Omega)$ and $\xi_{m}(\Omega)$
of order $m=S$. In the Gaussian model they are Gauss-Hermite function, therefore we get the analytic expressions
\begin{align}
\Delta_{\mathrm{s}}^2&=\left(S+\frac{1}{2}\right)\sigma_{\mathrm{s}}^{-2},\label{Deltas}
\\
\Delta_{\mathrm{i}}^2&=\left(S+\frac{1}{2}\right)\sigma_{\mathrm{i}}^{-2}.\label{Deltai}
\end{align}

\vspace{10pt}
\textit{Temporal resolution}: the Schmidt number and the eigenfunctions can be used for obtaining an analytic expression -- in the case of the Gaussian approximation -- 
of the resolution $r$ of the imaging scheme. 
The resolution of an imaging scheme is the smallest detail that can be transferred; hence, by following \cite{Bertero1982}, we can
estimate $r$ as the average distance, in time domain, of the zeros of the $S$-th eigenfunction of the impulse response function: the smaller $r$ is the better the resolution of the scheme.
In the general case of non-ideal phase matching, the transfer function $h_{\mathrm{i}}(\Omega,\Omega')$ is characterized by the two families 
of eigenfunctions $\{\xi_m\}$ and $\{\zeta_m\}$ (see \eqref{hi diag}). In time domain, the first family of eigenfunctions determine the characteristic time $r_{\mathrm{i}}$ of the
system in the image plane, while the second family determines the characteristic time $r_{\mathrm{s}}$ in the object plane.
Then the resolution of the system, at the image plane, is given by
\begin{align}
r=\mathrm{max}\left\{|M|r_{\mathrm{s}},r_{\mathrm{i}}\right\}.
\label{res}
\end{align}
where $r_{\mathrm{s}}$ and $r_{\mathrm{i}}$ are evaluated as the average distance of the zeros of the inverse Fourier transform of
$\zeta_{m}(\Omega)$ and $\xi_{m}(\Omega)$ for $m=S$. This distance is given by the temporal width of the eigenfunction
divided by the number of its semi-oscillations. By using the inverse Fourier transform of $\{\xi_m\}$ and $\{\zeta_m\}$, we find that the temporal widths~\cite{twidth} for the eigenfunctions $m=S$ are
\begin{align}
T_{\mathrm{s}}&=\sigma_{\mathrm{s}}
\sqrt{\left(S+\frac{1}{2}\right)\left(1+\frac{(D_{\mathrm{in}}-D_1)^2}{\sigma_{\mathrm{s}}^4}\right)}
\\
T_{\mathrm{i}}&=\sigma_{\mathrm{i}}
\sqrt{\left(S+\frac{1}{2}\right)\left(1+\frac{(D_{\mathrm{out}}-D_2)^2}{\sigma_{\mathrm{i}}^4}\right)}
\end{align}
with $D_1=D_{\mathrm{f}}(1-\widetilde{\Delta}_h^2/\Delta_{\mathrm{p}}^2)$ and $D_2=D_{\mathrm{f}}(1-\widetilde{\Delta}_v^2/\Delta_{\mathrm{p}}^2)$.
Notice that, in $T_{\mathrm{s}}$ and $T_{\mathrm{i}}$, the parameters $\sigma_{\mathrm{s}}$, $\sigma_{\mathrm{i}}$ and $S$ depend
on the details of the phase-matching profile of the SFG process, while the parameters $D_{\mathrm{in}}$ and $D_{\mathrm{out}}$ depend on the
imaging scheme and they are related each other via the imaging condition \eqref{imag cond}.

Since the number of semi-oscillations, for a $S$-th order Gauss-Hermite function, is given by $S$ then
\begin{align}
r_{\mathrm{s}}&=T_{\mathrm{s}}/S,
\label{rin}
\\
r_{\mathrm{i}}&=T_{\mathrm{i}}/S.
\label{rout}
\end{align}

\vspace{10pt}
\noindent
\textit{Temporal Field of View (FOV)}: this figure of merit is defined as the temporal duration over which an object can be viewed. By following \cite{Bennett2000b}, let's assume that the input signal modes are made up of short classical features $f_{\mathrm{in}}(\tau;\tau_0)$ centered at $\tau_0$ and that, at the output, they are transformed as $f_{\mathrm{out}}(\tau;\tau_0)$. Then the FOV is defined as the width of the energy profile $U(\tau_0)$ outgoing the system as a function of the input $\tau_0$ of this feature
\begin{equation}
U(\tau_0)\propto
\int_{-\infty}^{+\infty}\mathrm{d}\tau
|f_{\mathrm{out}}(\tau;\tau_0)|^2.
\label{energy out}
\end{equation}
The estimation of this quantity in the case of a non ideal phase-matching profile requires a numerical calculation. However the modal approach allows to simplify expression \eqref{energy out}.
Indeed by using \eqref{Ai out} and the classical part of \eqref{Aout decomp} we obtain
\begin{equation}
f_{\mathrm{out}}(\tau;\tau_0) = \sum_{m=0}^{+\infty} (-s_m) f_{\mathrm{in},m}(\tau_0)
\end{equation}
with
\begin{equation}
f_{\mathrm{in},m}(\tau_0)=
\int_{-\infty}^{+\infty}\mathrm{d}\tau\,
\zeta_k^*(\tau)f_{\mathrm{in}}(\tau;\tau_0),
\end{equation}
where $\zeta_k^*(\tau)$ is the Fourier transform of $\zeta_k^*(\Omega)$.
As a consequence, one has
\begin{equation}
U(\tau_0)\propto
\sum_{m=0}^{+\infty} s_m^2
|f_{\mathrm{in},m}(\tau_0)|^2.
\label{energy out II}
\end{equation}
Expression \eqref{energy out II} can be considered a generalization of expression (28) in \cite{Bennett2000b} for high conversion efficiency regimes and when the phase-matching is non ideal.
The FOV of the system can be estimated, then, by
considering that the width of $U(\tau_0)$ is given
by the width of the element $f_{\mathrm{in},m}(\tau_0)$ with $m=S$.
As an example, consider an object pulse having infinitely small details such that $f_{\mathrm{in}}(\tau;\tau_0)\rightarrow\delta(\tau-\tau_0)$. In this case $f_{\mathrm{in},m}(\tau_0)=\zeta^*_m(\tau_0)$ and the FOV approximately is given by the time duration of the input eigenvector corresponding to $m=S$, hence FOV$\approx T_{\mathrm{s}}$.

As we discussed in~\cite{Patera2017}, a time lens with a FOV designed for classical images is not necessary adapted for the manipulation of a quantum image because while the degradation of its classical part might be negligible at the same time it will be polluted by vacuum noise in a measure quantified by the $c_m$ coefficients. The modal approach allows us to choose the level of acceptable added noise by simply choosing the maximum allowed $c_{\mathrm{max}}$. From this choice, then, one can extract the order $M$ such that $c_{M}\le c_{\mathrm{max}}$. When $c_{\mathrm{max}}<50\%$ then $M<S$. Hence the quantum-FOV corresponding to the chosen level of added noise can be defined such as the time duration of 
\begin{equation}
\sum_{m=0}^{M} s_m^2
|f_{\mathrm{in},m}(\tau_0)|^2.
\end{equation}
\vspace{10pt}
\noindent
\textit{Time-bandwidth product}: an important figure of merit in (temporal) imaging is the time-bandwidth product of the time lens. It corresponds to the number of (temporal) features that can be processed by a (time) lens and it is given by the ratio $\text{FOV}/r$ \cite{Bennett2000b,Salem2013}.

In order to improve the system performances one has to increase the Schmidt number. This can be done by choosing
the experimental parameters such that the ratio $\rho=(F_1-F_2)/(F_1+F_2)$ is as close as
possible to one or equivalently $F_1\gg F_2$.
As an example in figure \ref{smcm} we compare two cases where the ratio $\rho$ is changed
from $0.8$ to $0.9$. This increment allows a doubling 
of $S$ (from $6$ to $13$). Notice, however, that increasing the Schmidt number of the
imaging scheme comes at a cost. Indeed for any choice, since optimal conversion efficiency is obtained
for $s_0=1$ or, equivalently, $g l_c \lambda_0 = \pi/2$, then increasing the value of $S$ means
increasing the value of $g l_c A_{\mathrm{p}}$.

\subsection{Ideal situation}
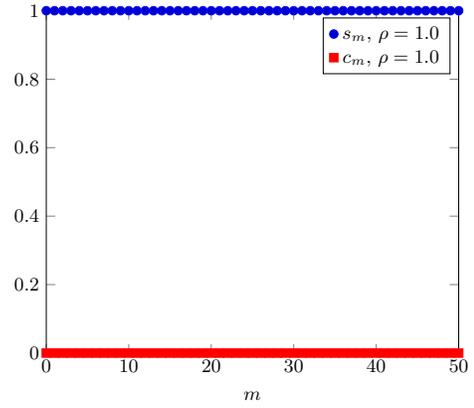
\begin{figure}[t!]
\begin{center}
\begin{tikzpicture}[scale=0.8]
\begin{axis}
[
xlabel={$m$},
ylabel={},
xmin=0, xmax=50,
ymin=0, ymax=1,
]
\addplot+[ycomb, only marks, domain=1:10, samples at={0,1,...,50}, blue]{sin(2*atan(1)*(1.)^(x/2))};
\addlegendentry{$s_m$, $\rho=1.0$}
\addplot[ycomb, only marks, mark=square*, red, domain=1:10, samples at={0,1,...,50}]{cos(2*atan(1)*(1.)^(x/2))};
\addlegendentry{$c_m$, $\rho=1.0$};
\end{axis}
\end{tikzpicture}
\end{center}
\caption{Plot of $s_m=\sin(\lambda_m)$ (blue-circles) and $c_m=\cos(\lambda_m)$ (red-squares) for 
$g l_c\lambda_0=\pi/2$ and $\rho=1$ (ideal configuration).}
\label{smcm_ideal}
\end{figure}
\begin{figure*}[!ht]
\centering
\subfigure[\label{fig:K1 module perfPM}]{\resizebox{0.45\textwidth}{!}{\includegraphics{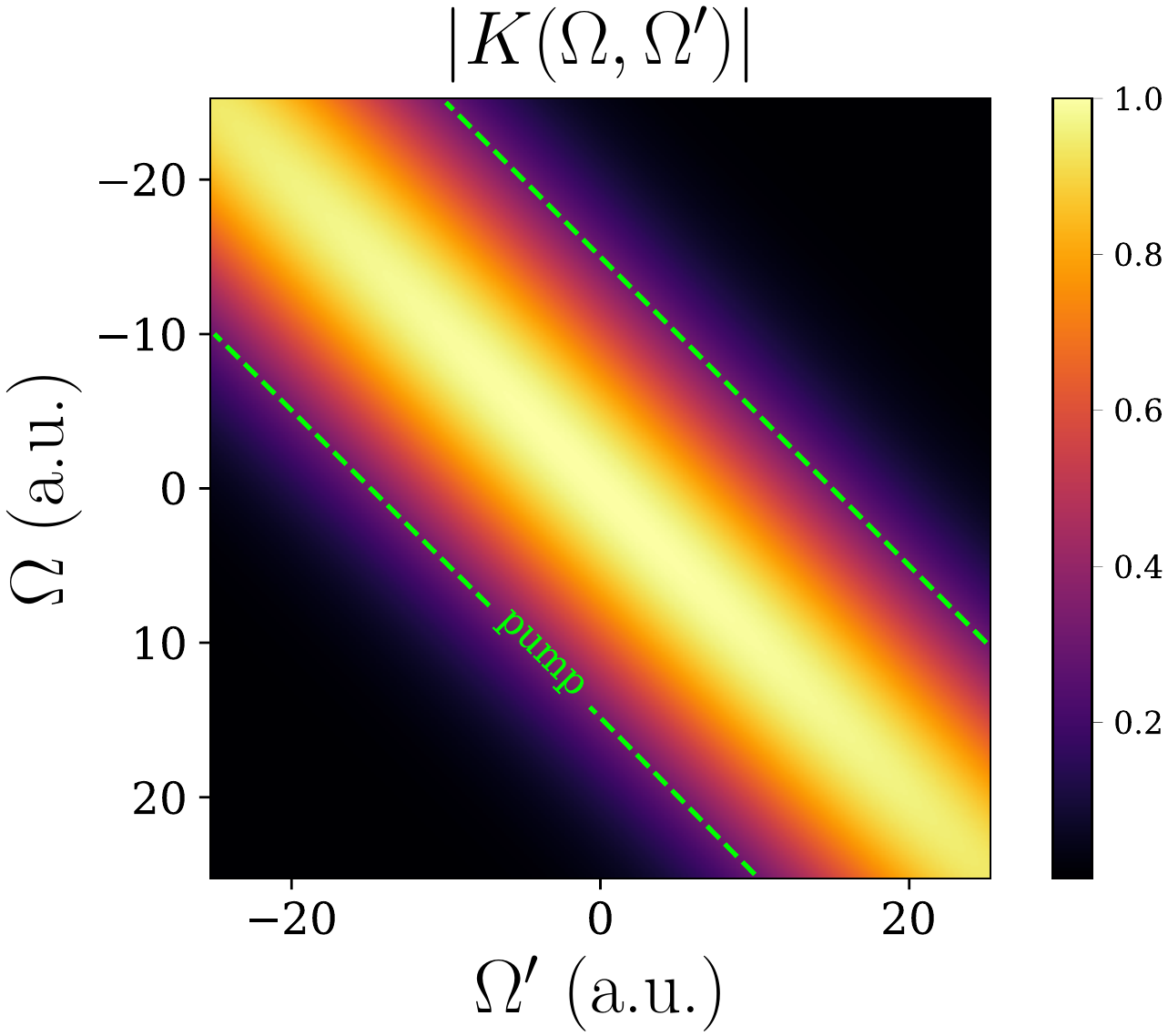}}}
\quad
\subfigure[\label{fig:K1 phase perfPM}]{\resizebox{0.45\textwidth}{!}{\includegraphics{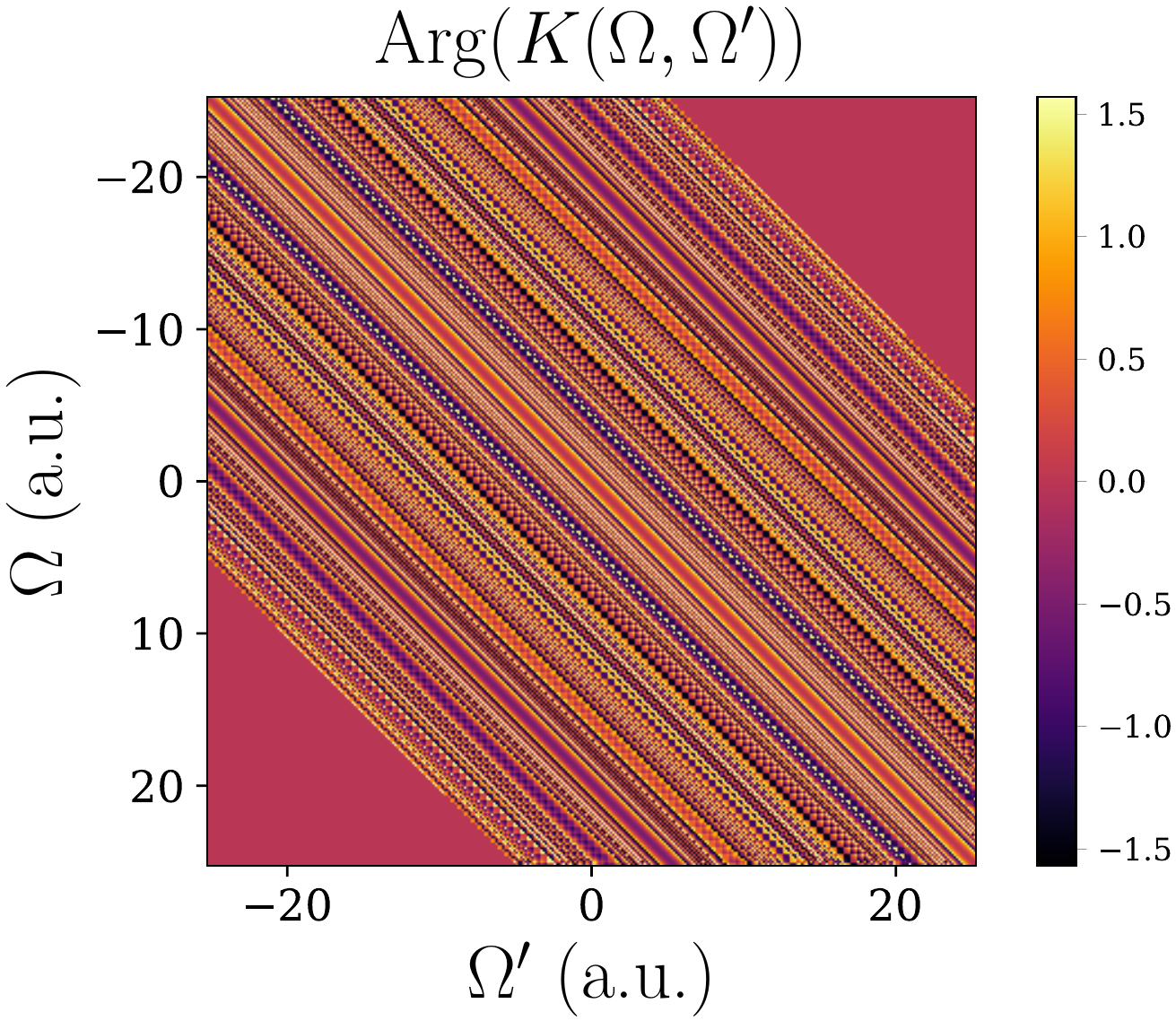}}} \\
\subfigure[\label{fig:eval perfPM}]{\resizebox{0.6\textwidth}{!}{\includegraphics{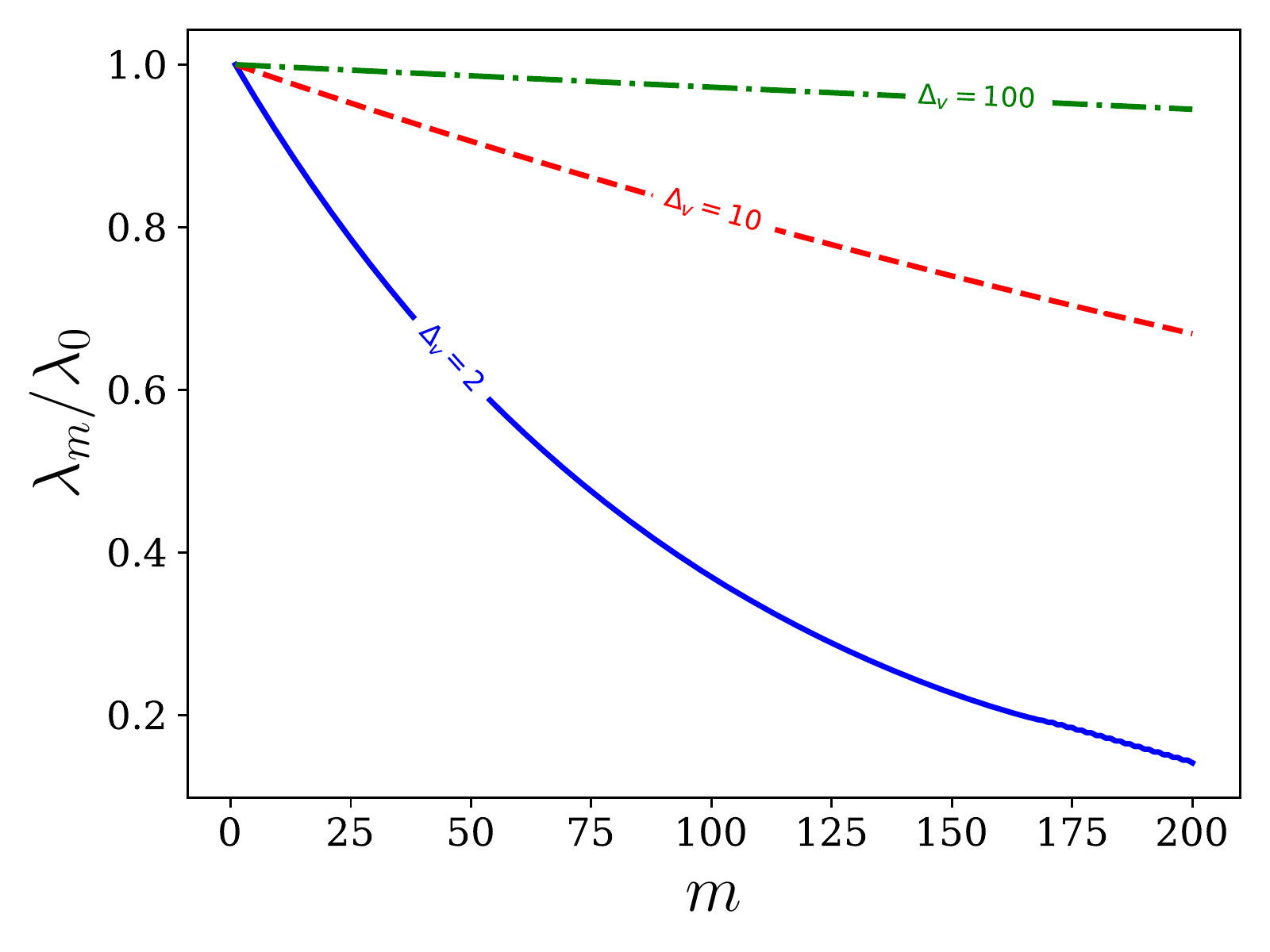}}}
\caption{(a) Module and (b) phase of $K(\Omega,\Omega')$ as described in eq. \eqref{K1} with parameters 
$\Delta_h=100$ a.u., $\Delta_v=100$ a.u., $\Delta_{\mathrm{p}}=10$ a.u., $D_{\mathrm{f}}=5$ a.u. 
(c) Comparison between the eigenspectrum of three configuration tending to the perfect phase-patching case: 
solid blue for $\Delta_v=2$ a.u., dashed red for $\Delta_v=10$ a.u. and dot-dashed green for $\Delta_v=100$ a.u. 
Notice that the case $\Delta_v=2$ a.u. corresponds to the one depicted in figure \ref{fig:K1}.}
\label{fig:K1 perfectPM}
\end{figure*}
The ideal situation (see configuration (i) in section II) is reached when the phase-matching is perfect over a very broad bandwidth 
$\Delta_h,\Delta_v \rightarrow+\infty$, when the pump has a very large duration
$\tau'_{\mathrm{p}}=D_{\mathrm{f}}\Delta_{\mathrm{p}}\rightarrow+\infty$ and 
when the condition \eqref{pi2bis} for perfect conversion efficiency is verified.
In this case $s_m\approx 1$ and $c_m\approx 0$ for all $m$ (see figure~ \ref{smcm_ideal}) so that, from eq. \eqref{Aout decomp}, no noise is introduced.
The Schmidt number $S\rightarrow+\infty$ and, consequently, the bandwidths $\Delta_{\mathrm{s}}$ and $\Delta_{\mathrm{s}}$ are arbitrarily large
and the parameters $r_{\mathrm{s}}$ and $r_{\mathrm{i}}$ are arbitrarily small.

The connection between the modal and the traditional approach to temporal imaging is easily obtained by observing that in this limit
\begin{align}
\sum_m \xi_m(\Omega)\zeta_m^*(\Omega') &= 
\mathcal{G}_{\mathrm{out}}(\Omega)\mathrm{e}^{\frac{\mathrm{i}}{2}D_{\mathrm{f}}(\Omega-\Omega')^2}\mathcal{G}_{\mathrm{in}}(\Omega').
\end{align}
When the imaging condition \eqref{imag cond} is satisfied, 
eq. \eqref{Ai out} takes the form of the well known unitary transformation for perfect
quantum temporal imaging \cite{Patera2015}
\begin{align}
\hat{A}_{\mathrm{out}}(\tau)=
-\frac{1}{\sqrt{M}}\mathrm{e}^{\mathrm{i}\frac{\tau^2}{2 M D_{\mathrm{f}}}}
\hat{A}_{\mathrm{in}}(\tau/M).
\end{align}
Notice that in this ideal situation $F_1\rightarrow+\infty$ and $F_2\rightarrow 0$. Therefore, since $A_{\mathrm{p}}\propto (F_1+F_2)^{1/2}$,
this regime would require the nonphysical situation of infinitely large pump amplitude $A_{\mathrm{p}}\rightarrow+\infty$, 
a condition that is required for perfectly up-converting all the infinite number of input modes.

\subsection{Perfect Phase-Matching and finite aperture}\label{perf PM finite aperture}

In this subsection we consider the case where the phase-matching is almost perfect ($\Delta_h,\Delta_v \gg\Delta_{\mathrm{p}}$)
but the imaging scheme presents a finite aperture induced by the pump pulse duration 
$\tau'_{\mathrm{p}}=D_{\mathrm{f}}\Delta_{\mathrm{p}}$ (see configuration (ii) in section II). Notice that in this case the traditional approach to temporal imaging (classical \cite{Kolner1994}
and quantum \cite{Patera2018}) gives analytic results as reviewed in Section \ref{standard QTI}. The purpose of this section is, then, to test the modal approach developed in this paper
by comparing its predictions to those obtained by the traditional approach. In figure \ref{fig:K1 perfectPM} we show the module and argument of 
$K(\Omega,\Omega')$ as well as a comparison between the eigenspectrum of two non ideal configurations as those in figure \ref{fig:K1} (blue solid and red dashed lines) and a configuration where the phase-matching tends to ideal (green dot-dashed line).

Here we have $\widetilde{\Delta}_h\approx\Delta_{\mathrm{p}}(1-\Delta_{\mathrm{p}}^2/\Delta_{h}^2)$ and
$\widetilde{\Delta}_v\approx\Delta_{\mathrm{p}}(1-\Delta_{\mathrm{p}}^2/\Delta_{v}^2)$. Since, typically, the pump pulse is dispersed in the Fraunhofer limit 
($D_{\mathrm{f}}\Delta_{\mathrm{p}}^2\gg1$), then $F_1\approx D_{\mathrm{f}}\Delta_{\mathrm{p}}^2$ and $F_2\approx\Delta_{\mathrm{p}}/\Delta_{hv}$, where
\begin{align}
    1/\Delta_{hv}^2&=1/\Delta_{h}^2+1/\Delta_{v}^2.
\end{align}
Also we have $\sigma_{\mathrm{i}}\approx\sqrt{D_{\mathrm{f}}\Delta_{\mathrm{p}}/\Delta_{hv}}$
and $\sigma_{\mathrm{s}}\approx\sqrt{D_{\mathrm{f}}\Delta_{\mathrm{p}}/\Delta_{hv}}$. The Schmidt number of the imaging scheme is large $S\gg1$,
a consequence of the fact that the spectrum of $s_m$ falls off very slowly. There is not an explicit form for the expression of $S$ in the general
case, however an underestimated value can be obtained in the low-gain regime (i.e. $g l_c \lambda_0\ll \pi/2$)
\begin{align}
S\approx \tau'_{\mathrm{p}} \Delta_{hv}.
\end{align}
By using this expression in \eqref{rin} and \eqref{rout}, we find that the $r_{\mathrm{s}}=(1+1/|M|)D_{\mathrm{f}}/\tau'_{\mathrm{p}}$ and 
$r_{\mathrm{i}}=(1+|M|)D_{\mathrm{f}}/\tau'_{\mathrm{p}}$. Hence in the limit of large magnification $|M|\gg1$ the resolution $r$ \eqref{res} is
\begin{align}
r \approx |M| D_{\mathrm{f}}/\tau'_{\mathrm{p}}
\label{perfectPM M large res}
\end{align}
and in the limit of large compression $|M|\ll 1$ 
\begin{align}
r \approx D_{\mathrm{f}}/\tau'_{\mathrm{p}}
\label{perfectPM small M res}
\end{align}
which corresponds to the resolution obtained in \cite{Kolner1994} and \cite{Patera2018}.

Connection with the traditional approach is obtained by observing that in this limit (see \eqref{Kbis})
\begin{align}
\sum_m \lambda_m \xi_m(\Omega)\zeta_m^*(\Omega') &\approx
\mathcal{G}_{\mathrm{out}}(\Omega)
\alpha_{\mathrm{p}}(\Omega-\Omega')
\mathcal{G}_{\mathrm{in}}(\Omega').
\end{align}
As a consequence the impulse response results to be the Fourier transform of the pupil function of the scheme as
described in \cite{Kolner1994} and \cite{Patera2018},
and in section \ref{standard QTI} of the present paper.

It is interesting to observe that, from the expressions of $S$, $T_{\mathrm{s}}$ and $T_{\mathrm{i}}$, the limit of $\Delta_h,\Delta_v \gg\Delta_{\mathrm{p}}$ gives arbitrary large values. However their ratios~\eqref{rin} and~\eqref{rout} remain finite as well as the system resolution.

\subsection{Finite phase-matching and finite aperture}
This is the most general case (see configuration (iii) in section II), we assume the typical situation where the group velocities of pump and signal field are
matched; in this case, hence, we have $\Delta_v\ll\Delta_{\mathrm{p}}\ll\Delta_h$. This situation corresponds to that depicted
in figure \ref{fig:K1}. In this case we have $\widetilde{\Delta}_{hv}\approx\Delta_v$, $\widetilde{\Delta}_v\approx\Delta_v$ and $\widetilde{\Delta}_h\approx\Delta_{\mathrm{p}}$. This implies that $F_1\approx\tau'_{\mathrm{p}}\Delta_v$ and $F_2\approx 1-(\Delta_v/\sqrt{2}\Delta_{\mathrm{p}})^2$. Then the Schmidt number $S\approx\tau'_{\mathrm{p}}\Delta_v$ is smaller than the situation with perfect phase-matching in section \ref{perf PM finite aperture} by a factor $\Delta_{\mathrm{v}}/\Delta_{\mathrm{p}}$. Also we have that $\sigma_{\mathrm{s}}\approx\sqrt{D_{\mathrm{f}}\Delta_v/\Delta_{\mathrm{p}}}$ and $\sigma_{\mathrm{i}}\approx\sqrt{D_{\mathrm{f}}\Delta_{\mathrm{p}}/\Delta_v}$. 

For large magnification $|M|\gg1$ the temporal duration of the eigenmodes $\xi_m(\tau)$ and $\zeta_m(\tau)$ with $m=S$ are $T_{\mathrm{s}}\approx\tau'_{\mathrm{p}}$ and $T_{\mathrm{i}}\approx\tau'_{\mathrm{p}}\Delta_{\mathrm{p}}/\Delta_v$. In this case the resolution of the system, evaluated from \eqref{res}, is
\begin{equation}
r\approx\mathrm{max}\left\{|M|r_{\mathrm{s}},\frac{\Delta_{\mathrm{p}}}{\Delta_v}r_{\mathrm{s}}\right\}
\end{equation}
with $r_{\mathrm{s}}\approx D_{\mathrm{f}}/\tau'_{\mathrm{p}}$.
As a consequence the resolution will depend on the relative magnitude between the magnification and the ratio $\Delta_{\mathrm{p}}/\Delta_v$. By comparison with \eqref{perfectPM M large res}, we see that if this ratio $\Delta_{\mathrm{p}}/\Delta_v$ is smaller than the magnification, then the effects of bandwidth filtering induced by the temporal walk-off between the pump and idler waves do not affect the resolution that is equal to the resolution of the case with perfect phase-matching. On the contrary, when the ratio $\Delta_{\mathrm{p}}/\Delta_v$ is larger than the magnification, the resolution is worst than that of the case of \eqref{perfectPM M large res} by an amount of $\Delta_{\mathrm{p}}/|M|\Delta_v$.

On the other side, for a system designed for a large compression $|M|\ll1$ the resolution results to be
\begin{equation}
r\approx\frac{D_{\mathrm{f}}}{\tau'_{\mathrm{p}}}\,
\frac{\Delta_{\mathrm{p}}}{\Delta_v}.
\end{equation}
In this case, by comparison with \eqref{perfectPM small M res}, the resolution is always worst than that obtained for the perfect phase-matching case by a factor of $\Delta_{\mathrm{p}}/\Delta_v$.

\section{Conclusions}
In this work we developed the modal approach for a SFG-based quantum temporal imaging scheme in the high conversion efficiency regime and for the general case of non-perfect phase matching and finite temporal aperture. In general this problem does not admit a closed-form expression for the impulse response function and for its Fourier transform, the transfer function. However, by using the modal approach, we showed that it is possible to express the transfer function in terms of its expansion on the singular values and eigenmodes of the problem. Then we showed how to obtain the relevant figures of merit of the imaging scheme and how to express them in terms of the modal decomposition. This allows to assess the performances of the QTI scheme. We finally used these results for comparing the relevant figures of merit to those obtained in the regime of prefect phase-matching and infinite aperture and of perfect phase-matching and finite aperture. This comparison makes clear the necessity of a multimode operation for implementing a QTI scheme working in high conversion efficiency regimes and the physical parameters that need to be adjusted in order to improve its performances. Our results will allow, therefore, for better designs for noiseless manipulation of the spectrotemporal degrees of freedom of photonic non-classical states.
\section{Acknowledgments}
This work was supported by the network QuantERA of the European Union’s Horizon 2020 research and innovation programme under project “Quantum information and communication with high-dimensional encoding” (QuICHE).


\begin{thebibliography}{99}
\bibitem{twidth} $T_{\mathrm{s}}$ and $T_{\mathrm{i}}$ are evaluated in terms of the modes variances.
\end{thebibliography}

\end{document}